\documentclass{JHEP3}
\usepackage{amsmath,amsfonts,bbm,euscript, amssymb}
\usepackage{epsfig}
\usepackage{graphicx}
\usepackage{psfrag}

\numberwithin{equation}{section}

\def\be{\begin{equation}}
\def\ee{\end{equation}}
\def\baray{\begin{eqnarray}}
\def\earay{\end{eqnarray}}
\def\ba{\begin{eqnarray}}
\def\ea{\end{eqnarray}}

\title{Phenomenology of Infrared Smooth Warped Extra Dimensions}
                                                                                
\author{Paul McGuirk, Gary Shiu, Kathryn M. Zurek \\
Department of Physics, University of Wisconsin Madison, 
WI 53706, USA \\ E-mail: \email{mcguirk@wisc.edu},
\email{shiu@physics.wisc.edu}, \email{kzurek@wisc.edu}}

\date{\today}

\abstract{We study the effect of the infrared (IR) geometry on the phenomenology of warped extra dimensions with gauge and fermion fields in the bulk.   We focus in particular on a ``mass gap'' metric which is AdS in the ultraviolet, but asymptotes to flat space in the IR, breaking conformal symmetry.  These metrics can be dialed to approximate well the geometries arising in certain classes of warped string compactifications.  We find, similar to our earlier results on the Kaluza-Klein (KK) graviton, that these metrics give rise to phenomenologically significant shifts in the separation of KK gauge modes in the mass spectrum (up to factors $\sim 2$) and their couplings to IR localized fields (up to factors $\sim 5 - 10$ increase).   We find that, despite shifts in the spectra, the constraint $m_{KK} \gtrsim 3 \mbox{ TeV}$ from $S$ remains robust in the class of 5-d mass gap metrics, and that the change to $T$ is not significant enough to remove the need for custodial symmetry.}
                                                                                
\preprint{MAD-TH-07-12, MADPH-07-1501}

\begin{document}

\bibliographystyle{JHEP}

\section{Introduction}\label{sec:intro}

Extra dimensions appear in many extensions of the Standard Model (SM), from solutions to the hierarchy problem, to efforts to unify gravity with the rest of the forces in string theory.  One appealing solution to the hierarchy problem comes from warped extra dimensions, where Randall and Sundrum (RS) \cite{Randall} showed that the 5-dimensional Anti-de Sitter (AdS) space appearing often in string theories can produce a naturally small Higgs boson mass through exponential suppression of the Planck scale.

Since the idea of RS appeared almost a decade ago, much work has been done in the phenomenology of warped extra dimensions.  Much of this work relates to embedding the SM in the extra dimension, and the resultant implications for collider searches (e.g. \cite{Lillie1, Lillie2, AgasheEWGauge, AgasheGraviton, Agashe}).    While allowing the SM fields to propagate in the bulk of the extra dimension alleviates problems with Flavor Changing Neutral Currents \cite{Huber1, Gherghetta} and allows an understanding of the flavor hierarchy, other problems and constraints are introduced, most notably from the Peskin-Takeuchi parameters, $S$ and $T$ \cite{Csaki1, Carena1, Carena3, Carena4}, and from $B$ factories \cite{Agashe, AgasheFlavor2}.   With additional work (usually involving the addition of other symmetries, such as custodial symmetry in the AdS bulk \cite{AgasheCustodial}, or an RS GIM mechanism \cite{RSGIM}),  the problems can be alleviated and constraints relaxed, making Kaluza-Klein (KK) gauge bosons with $m_{KK} \gtrsim 3 \mbox{ TeV}$ consistent with current data, and within observational reach of the Large Hadron Collider (LHC).

All of these analyses, both collider studies and constraints from precision Electroweak (EW), were carried out within the context of  pure AdS geometry.
The AdS metric, however, is singular in the IR.
Moreover, the hierarchy is not stabilized as the warp factor can be arbitrarily small. 
An attempt to stabilize the weak-Planck hierarchy in the RS scenario has 
been made in \cite{Goldberger:1999uk}, by introducing a bulk field that generates a potential for the distance between the IR and UV branes.  However,
it remains to be seen if this mechanism is realized in a fully UV complete model. Furthermore,
as one allows fermion and gauge fields to propagate in the bulk, the motivation for introducing
the IR and UV branes becomes less clear.
Interestingly, warped models can be constructed in string theory where the IR geometry is smooth.
 As a result, the hierarchy is stabilized by the smooth IR geometry without the need of introducing an 
 IR brane.
 A particularly well studied example of such warped geometry
 is the Klebanov-Strassler (KS) throat  \cite{KS}. The KS geometry is non-compact, but upon compactification with fluxes (see e.g. \cite{fluxreview} and
examples therein), the weak scale hierarchy is stabilized by quantized fluxes:
 \begin{equation}
 m_{TeV} = M_{Pl} e^{A_{min}} = M_{Pl} e^{-2 \pi K/3M g_s} 
 \end{equation}
 where $K$ and $M$ are 3-form flux quanta, $g_s$ is the string coupling, and $e^{A_{min}}$ is the minimum (non-vanishing) warp factor in the IR.  
In addition to 
the KS geometry (and its generalizations to the baryonic branch
\cite{Butti:2004pk}) whose metrics are known,
a wide class of string theory backgrounds \cite{Cvetic:2005ft,Martelli:2005wy}
admit similar IR modifications although explicit metrics for 
 such smooth warped geometries are yet to be constructed. 
These warped throat solutions  are 10-d relatives of the 5-d RS metric which, instead of continuing to become exponentially small, go to a constant in the IR.  The other five dimensions
are angular coordinates (whose topology is 
$S^2 \times S^3$ in the examples above). 
Upon integrating out the angular KK modes (whose masses are higher than the low lying radial KK modes, see, e.g., \cite{Firouzjahi:2005qs}), one obtains an effective 5-d RS-like model.

In an earlier paper \cite{us}, we showed that these 10-d relatives of RS, though their metric differs only a little from the RS metric, and only in the last decade of the 16 decade hierarchy between the Planck and TeV scales, lead to dramatically different collider signatures.  In particular, we found that the spacing between the KK graviton modes with the KS metric changes by a factor of several relative to the KS case.  Their couplings to TeV localized states also changes significantly: the couplings of all KK graviton modes is no longer universal, and is larger by a factor of at least $\sim 30$.  We thus showed that the spectrum of KK graviton states can be a sensitive probe of the geometry of the warped background,  so that we may learn something about the nature of warped string compactifications at the LHC (the effects of other types of geometries have also been studied in, e.g. \cite{Kaloper:2000jb,Dienes:2001wu}).

The KS metric is complicated, and thus presents a barrier to doing detailed phenomenological model building.  It has been shown, however, that the KS metric (and its generalizations mentioned above) can be well approximated by a simple ``mass gap'' metric \cite{Kecskemeti:2006cg,us}, which interpolates between AdS warping in the UV and flat space in the IR.
In the AdS/CFT language, the ``mass gap'' metric corresponds to a breaking of the conformality in the IR, 
with the mass gap parameter set by the confinement scale of the field theory dual.  Such modifications have been shown to have important effects in the contexts
of inflation~\cite{Shiu:2006kj} or phase transitions~\cite{Hassanain:2007js}.
In our earlier paper~\cite{us}, we also numerically solved for the masses and wavefunctions of the KK gravitons with the mass gap metric in 5-d and 10-d and again found large differences in the masses and wavefunction overlaps with the IR brane.  If more than one KK graviton can be observed at the LHC, the ratio of the masses and ratio of couplings to SM states measures the background metric, and may give insight into the string compactification which generates the warping.

Here we extend our previous work to studies of gauge bosons propagating in the bulk.  We restrict ourselves to mass gap metrics--they may be handled numerically much more easily than KS metrics, and can be dialed to reproduce 
the desired 
IR behavior.  We solve the spectrum with both the 5-d and 10-d mass gap metrics, but restrict ourselves to five dimensions
in the precision EW analysis on $S$ and $T$ in order to compare 
our results with that of the well-studied
RS models.
Furthermore, the Higgs mechanism 
cannot be embedded in the way that it is in the extended RS scenario
into higher-dimension
mass gap metrics  without
placing restrictions on the angular space.
While much progress has been made toward 
constructing realistic
D-brane models \cite{Blumenhagen:2005mu,Blumenhagen:2006ci,Marchesano:2007de},
explicit realistic warped string models with features of the extended RS scenario have yet to be found. The present work is thus a first step in exploring the electroweak constraints of such warped string models. Our results could be adopted to specific compactifications once this scenario finds its embedding in string theory. 

We are interested in two primary questions.  First, how do the spectrum and couplings of KK gauge bosons change with the new metric?  Can we learn about how conformality is broken in the IR from the spectrum of KK states?  This is a natural follow-up to the earlier study on KK gravitons, which showed strong sensitivity to the IR behavior of the metric.  Second, very detailed model building and precision EW constraints have been derived in the context of the AdS background.  Do these constraints change much when a new metric is introduced?  If the KK spectrum changes significantly with each new metric, the constraints on the lowest KK gauge boson mass from $S$ (currently sitting at $3 \mbox{ TeV}$) may be modified (this question was also considered in \cite{Delgado}).  Even a modest shift in the constraint from 3 TeV to 1 TeV, for example, will change LHC physics reach dramatically. 

The outline of this paper is as follows.  In the next section we set up the equations for gauge and fermions propagating in the bulk, with an arbitrary number of extra dimensions
with arbitrary modifications
and of the metric in the IR, numerically solving the equations to obtain the spectrum of states and couplings to IR localized fields. We apply the formalism to the case where the Higgs vev on the IR brane modifies the profiles of the gauge boson zero modes in the extra dimension.   We then turn to analyzing the
Peskin-Takeuchi parameters with the new metrics in 5-d and again compare the results and constraints to RS.  We consider the effect of the new metrics on the fermion profiles, including constraints on $Z \rightarrow b \bar{b}$.  We then conclude.

\section{Warped geometries with IR smooth behavior}

We consider a class of warped $D=\left(5+\delta\right)$-dimensional backgrounds
with line element
\begin{equation}
  \label{eq:mgmetric}
  ds^{2}=G_{MN}dx^{M}dx^{N}
  = f^{-1/2}\left(r\right)\eta_{\mu\nu}dx^{\mu}dx^{\nu}
  -f^{1/2}\left(r\right)\left(dr^{2} + r^{2}ds_{X^{\delta}}^{2}\right)
\end{equation}
where $\mu,\nu$ run over $0,1,2,3$ and $M,N$ run over all
dimensions.  The line element for the compact angular space $X^{\delta}$ is
\begin{equation}
  r^{2}ds_{X^{\delta}}^{2}=r^{2}\tilde{g}_{\xi\zeta}dy^{\xi}dy^{\zeta}
\end{equation}
where $\tilde{g}$ is independent of $x^{\mu}$ but not necessarily of $r$.
The determinant of the full metric tensor is then
$G=f^{\left(\delta-3\right)/2}r^{2\delta}\tilde{g}$.
In the Randall-Sundrum scenario, the warp
factor is given in terms of the AdS curvature $k = 1/R$ by
\begin{equation}
  f\left(r\right)=\frac{R^{4}}{r^{4}}.
\end{equation}
 The identification with the RS metric may be made by a substitution
$r = e^{-k y}$.  The location of the UV brane in the extra dimension is taken
to be $r=R$ where the warp factor becomes 1 and the warped throat ends.
In RS,
an IR brane is placed at $r=r_{\mathrm{tip}}$ where
$r_{\mathrm{tip}}/R=\epsilon=e^{-11.27\pi}=4.2\times 10^{-16}$
to address the hierarchy problem.

Backgrounds with smooth behavior in the
infrared can be parametrized using the mass gap ansatz \cite{Kecskemeti:2006cg,us}:
\begin{equation}
  f\left(r\right) = \frac{R^{4}}{r_{\mathrm{tip}}^{4} + f_{2}R^{2}r^{2} +
    r^{4}}.
\end{equation}
Note that unlike the RS metric, this warp factor remains finite as $r\to 0$.
This form of the warp factor is chosen so that a brane placed at $r=0$ will 
produce the same
hierarchy as the RS background while for $r\sim R\gg \sqrt{f_{2}R}$ the
space is approximately AdS.  This is the scenario that we consider.  As
discussed above, the IR brane is introduced to localize the Higgs\footnote{One can also localize the Higgs using instantons as in \cite{Acharya:2006mx}, in which case the IR brane can be eliminated.}
but is
not necessary to cut off the space or stabilize the hierarchy.

Over most of the 16 orders of magnitude of the Planck-Higgs hierarchy, the warping is very well approximated by AdS, differing from it only in the last order of magnitude before reaching the tip of the throat at $r = 0$ where the IR brane is located.
We plot the warp factor for several different choices of $f_2$ in
Fig.~\ref{fig:warpplot}.

\EPSFIGURE[ht]{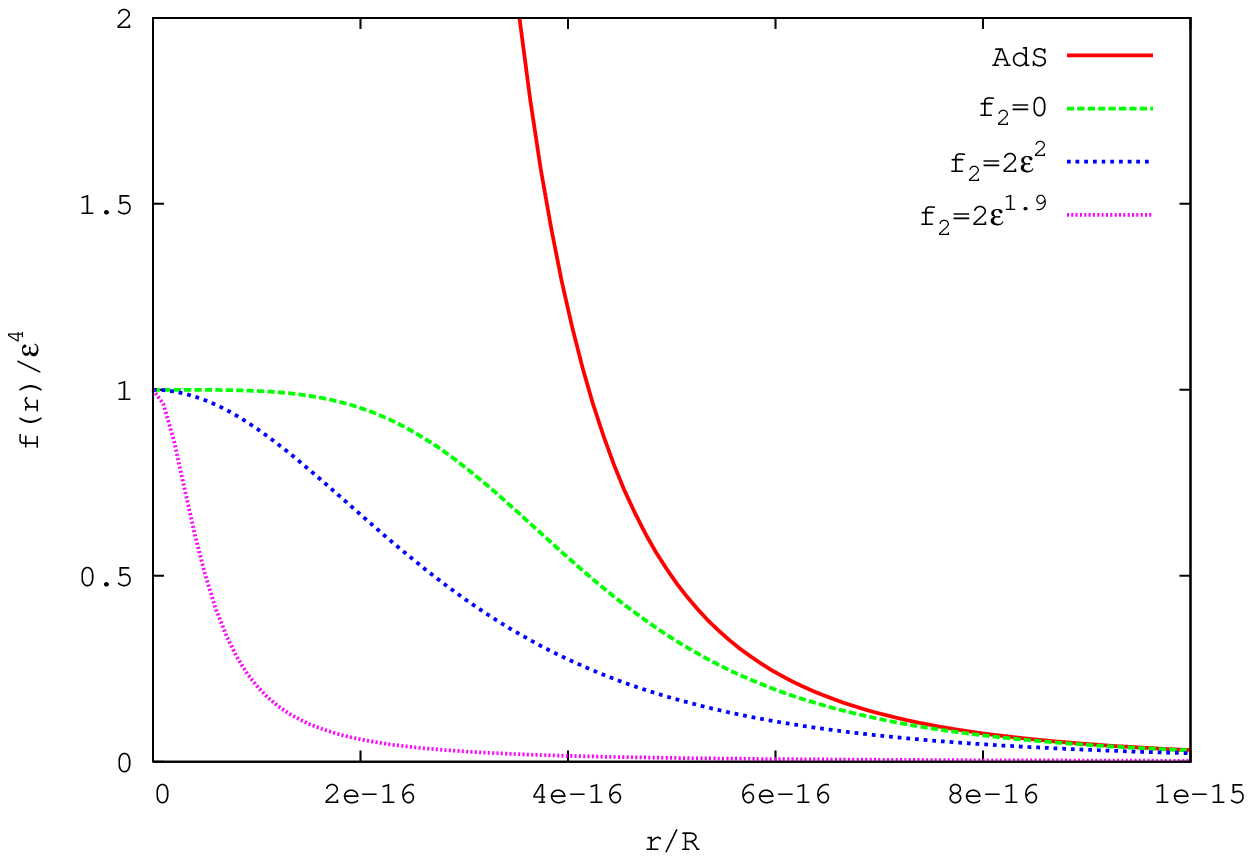,scale=.9}{The warp factor
for different geometries.  The warp factor for AdS grows without
bound while those for the mass gap geometries approach a finite value.  Thus the AdS space, unlike the mass gap metrics which are the subject of this paper, must be truncated at finite $r$.
The distinct shapes manifest themselves in many ways discussed here.
\label{fig:warpplot}}

\subsection{Gauge fields with 5-d and 10-d mass gaps}
\label{sect:gauge}

The physics for a gauge field propagating in the bulk of AdS was
considered in\cite{Davoudiasl:1999tf,Pomarol:1999ad}. We generalize the
analysis here for an arbitrary warp factor.  The action for a
canonically normalized gauge field $A_{M}$ propagating in the bulk of a
geometry described by Eq.~\ref{eq:mgmetric} is
\begin{equation}
  S = -\frac{1}{4}\int d^{5+\delta}x\,\sqrt{G}\,G^{MS}G^{NT}F_{MN}
  F_{ST} \notag \\
\end{equation}
where $F_{MN}=\partial_{M}A_{N}-\partial_{N}A_{M}$ is the field strength tensor
and the integral runs from the IR brane to the UV brane and over 
all of $X^{\delta}$.
The action is
\begin{equation}
  S= \int d^{5+\delta}x f^{\left(\delta-3\right)/4}r^{\delta}\sqrt{\tilde{g}}
  \left(-\frac{f}{4}
    \eta^{\mu\sigma}\eta^{\nu\tau}F_{\mu\nu}F_{\sigma\tau}
    + \frac{1}{2}\eta^{\mu\sigma}F_{\mu r}F_{\sigma r}
    + \frac{1}{2}\eta^{\mu\sigma}\frac{\tilde{g}^{\xi\zeta}}{r}
    F_{\mu\xi}F_{\sigma\zeta}\right) + ...,
\end{equation}
where the ellipses denote terms which involve only the fields from the six extra dimensions.  The mixing terms between our four dimensions and the six extra dimensions can be removed by an appropriate gauge choice.  The Kaluza-Klein decomposition of $A_{\mu}$ is then written
\begin{equation}
  A_{\mu}\left(x,r,y\right)=\sum_{n=0}^{\infty}\sum_{\left\{\ell\right\}}
  A_{\mu}^{\left(n,\left\{\ell
      \right\}\right)}
  \left(x\right)a_{n,\left\{\ell\right\}}\left(r\right)\Theta_{\left\{\ell\right\}}
  \bigl(y\bigr)
\end{equation}
where the set $\left\{\ell\right\}$ labels the angular mode.  If we limit
ourselves to the lighter $s$-wave modes (that is, where
$\Theta_{\left\{\ell\right\}}
  \bigl(y\bigr)$ is a constant) then the KK decomposition simplifies to
\begin{equation}
  A_{\mu}\left(x,r\right)=\sum_{n=0}^{\infty}A_{\mu}^{\left(n\right)}
  \left(x\right)a_{n}\left(r\right).
\end{equation}
After integrating over $r$, the KK fields
$A_{\mu}^{\left(n\right)}$ will be canonically normalized in the 4-d language
if we impose the orthonormality condition
\begin{equation}
  \label{eq:vnorm}
  \int dr\int d^{\delta}y\, f^{\left(\delta +1\right)/4}r^{\delta}\sqrt{\tilde{g}}\,
  a_{n}a_{m}=\delta_{nm}.
\end{equation}
The equation of motion will turn out to be an eigenvalue equation so that we can
impose this orthonormality condition so long as $a_{n}$ and
$a_{m}$ satisfy the same set of boundary conditions.

The $F_{\mu r}F_{\nu r}$ term in the $\left(5+\delta\right)$-d action corresponds
to mass terms for the KK
modes in the dimensionally reduced theory.  That is, after integrating by
parts, we can write the action as (again,
neglecting the effects of heavier angular modes)
\begin{equation}
  S = \int d^{4}x\sum_{n=0}^{\infty}\left\{-\frac{1}{4}\eta^{\mu\sigma}\eta^{\nu\tau}
    F^{\left(n\right)}_{\mu\nu}F^{\left(n\right)}_{\sigma\tau}
    + \frac{1}{2}m_{n}^{2}\eta^{\mu\sigma}A_{\mu}^{\left(n\right)}
    A_{\sigma}^{\left(n\right)}\right\},
\end{equation}
where the equation of motion for the radial wavefunction is
\begin{equation}
  \label{eq:vkkeom}
  \partial_{r}\left(f^{\left(\delta-3\right)/4}r^{\delta}\sqrt{\tilde{g}}\,
    \partial_{r}a_{n}\right) + f^{\left(\delta+1\right)/4}
  r^{\delta}\sqrt{\tilde{g}}\,m_{n}^{2}a_{n}=0.
\end{equation}
In order to write the action in this form, the boundary term resulting
from integration by parts must be made to vanish.  That is,
\begin{equation}
  \left[V_{X^{\delta}}f^{\left(\delta-3\right)/4}a_{m}\partial_{r}a_{n}
  \right]_{r=r_{\mathrm{IR}}}^{r=r_\mathrm{UV}}=0,
\end{equation}
where
\begin{equation}
  V_{X^{\delta}}=r^{\delta}\int d^{\delta}y\sqrt{\tilde{g}}
\end{equation}
is the (unwarped) volume of the angular space $X^{\delta}$.  The UV
contribution to the
surface term
can be made to vanish by imposing either the Neumann boundary condition
\begin{equation}
  \bigl.\partial_{r}a_{n}\bigr\vert_{r=r_{\mathrm{UV}}}=0
\end{equation}
or the Dirichlet boundary condition
\begin{equation}
  a_{n}\left(r=r_{\mathrm{UV}}\right)=0.
\end{equation}
The Dirichlet condition breaks 4-d gauge invariance so the Neumann condition
will be applied on the UV brane below.

If $\delta=0$ (i.e. 5-d spacetime) then a similar
argument can be applied to the boundary conditions on the IR.  However, if
$\delta\neq 0$, then the boundary conditions that can be consistently applied
depend on the shape of $X^{\delta}$.  If $X^{\delta}$ is such that
the volume shrinks to zero at the IR, then the requirement that the
boundary term vanishes is not
enough to determine the boundary condition.  For example, in the simplest
case, $X^{\delta}$ is a $\delta$-sphere, $S^{\delta}$, and $\tilde{g}$ is
independent of $r$.  In the near tip region, the warp function approaches
a constant $\left(R/r_{\mathrm{tip}}\right)^{4}$ and the equation of motion
Eq.~\ref{eq:vkkeom} simplifies to
\begin{equation}
  \partial_{r}\left(r^{\delta}\partial_{r}a_{n}\right)
  + \rho^{2}r^{\delta}a_{n}=0,
\end{equation}
where $\rho^{2}=m_{n}^{2}\left(R/r_{\mathrm{tip}}\right)^{4}$.  In 5-d, the
solutions
are sines and cosines.  In higher dimensions, the solutions in
the near-tip region can be written in terms of Bessel functions
\begin{equation}
  a_{n}\left(r\right)\to \frac{1}{r^{\left(\delta-1\right)/2}}
  \left(a J_{\left(\delta-1\right)/2}\bigl(\rho r\bigr)
    + b Y_{\left(\delta-1\right)/2}\bigl(\rho r\bigr)\right).
\end{equation}
The first term satisfies the Neumann condition $a'_{n}\left(r=0\right)=0$
while the second term diverges as $r\to 0$.  This means that in this case
the only boundary condition that be consistently applied is the Neumann
condition.  However, if the volume does not shrink to zero, then other boundary
conditions can be applied.

When the background is AdS, the arguments change.  If the angular space
is $S^{\delta}$, the effect of the $\delta$-dimensions drops out of the equation
of motion since the additional factor of $r^{\delta}$ is cancelled out by
the additional power of the warp factor.  This means that, up to normalization,
 the radial
wave function is independent of the existence of additional dimensions
(though there is still interesting physics when extra dimensions are added, as
in \cite{Davoudiasl:2002wz}).  Because of this independence, a divergent
solution exists in RS for any $\delta \ge 0$.  However, the space is cut off in
the Randall-Sundrum scenario by the IR brane before this divergence occurs.
 
As in the original RS scenario, we consider the Higgs to be localized on an IR
3-brane located at
at $r=r_{\mathrm{IR}}$ where $r_{\mathrm{IR}}=0$ for the mass gap geometries
while $r_{\mathrm{IR}}=r_{\mathrm{tip}}$ in RS.
In $5+\delta$ dimensions, the angular coordinates are taken to be
$y^{\xi}=0$.
If a gauge field propagating in the bulk couples to this Higgs with
coupling constant $g_{D}$, then after the
Higgs obtains a
vev $\tilde{v}$, the action for the gauge field can be written in
$\left(5+\delta\right)$-d language as
\begin{equation}
  S = \int d^{5+\delta}x\sqrt{G}\left\{-\frac{1}{4}G^{MS}G^{NT}F_{MN}F_{ST}
    + \frac{\tilde{v}^{2}g_{D}^{2}}{8}\frac{\delta\left(r-r_{\mathrm{IR}}\right)
      \delta^{\delta}\left(y\right)}{\sqrt{G_{rr}}r^{\delta}
      \sqrt{\tilde{g}}}G^{MS}A_{M}A_{S}\right\}
\end{equation}
where the determinant is from the Jacobian used in
transforming the
delta function from Cartesian coordinates to warped coordinates.
The equation of
motion for the internal wavefunctions does not change in the presence
of the Higgs vev, but the boundary
condition on the IR brane becomes modified \cite{Huber2}.  For
the geometries considered here, the boundary condition at the IR brane
becomes
\begin{equation}
  \bigl.\partial_{r}a_{n}\bigr\vert_{r=r_{\mathrm{IR}}}
  = \frac{\tilde{v}^{2}g_{D}^{2}}{4}\frac{\left
      [f\left(r=r_{\mathrm{IR}}\right)\right]^{\left(1-\delta\right)/4}}
  {V_{X^{\delta}}}a_{n}.
\end{equation}
For $\delta>0$, if the $X^{\delta}$ vanishes at the tip, then the boundary
condition is not well-defined.

As shown below and in~\cite{us}, increasing the number of dimensions in
an infrared smooth geometry has significant effects on the couplings and
spectra.  However, 
consistent
 localization of the Higgs on a TeV brane
depends on the angular geometry.  For this reason, we will consider
only 5-d backgrounds when doing the precision electroweak analysis below and leave the effects of higher-dimensional
backgrounds on $S$ and $T$ for future work.

In 5-d, the IR boundary condition becomes
\begin{equation}
  \partial_{r}a_{n} = \frac{\tilde{v}^{2}g_{5}^{2}}{4}
  \left[f\left(r=r_{\mathrm{IR}}\right)\right]^{1/4}a_{n}.
\end{equation}
With these boundary conditions, a non-vanishing constant solution (zero mode)
is not present.  Instead, the flat solution is replaced by an ``almost'' zero
mode where the profile is flat everywhere except near the IR
(Fig.~\ref{zprofile}).  This is also forces the 4-d mass to be non-vanishing
for the lowest state.

\subsection{Fermions with 5-d mass gaps}

Next we turn to generalizing bulk fermions for arbitrary warp factors.
This was done for the RS scenario in the context of neutrino masses in
\cite{Grossman:1999ra} though our conventions follow more closely those in
\cite{Altendorfer:2000rr}. We will specialize to 5-d because of the difficulties
with embedding the Higgs discussed above.
The action for a 5-d Dirac spinor $\Psi$ propagating in the bulk is
\begin{equation}
  S = \int d^{5}x\,\sqrt{G}\left\{i\bar{\Psi}\Gamma^{A}E^{M}_{\phantom{M}A}
    D_{M}\Psi - M_{f}\bar{\Psi}\Psi\right\},
\end{equation}
where $E^{M}_{\phantom{M}A}$ is the inverse of the f\"unfbein which is given in
this background by
\begin{equation}
  E_{M}^{\phantom{M}A} = \begin{pmatrix} f^{-1/4}\,\delta_{\mu}^{a} & 0\\
    0 & f^{1/4}\end{pmatrix},
\end{equation}
and $A,B,a,b$ are tangent space indices with $A,B$ ranging from $0$ to $4$
and $a,b$ ranging from $0$ to $3$.  We also denote by $\hat{M}$ the tangent
space index corresponding to the base space index $M$.
The covariant derivative is
\begin{equation}
  D_{M} = \partial_{M} + \frac{1}{4}\omega_{M}^{\phantom{M}AB}\Gamma_{AB},
\end{equation}
where $\omega_{M}^{\phantom{M}AB}$ is the spin connection and
$\Gamma_{AB}=\frac{1}{2}\left[\Gamma_{A},\Gamma_{B}\right]$ generate Lorentz
transformations.  We use the
representation $\Gamma_{\hat{\mu}}=\gamma_{\mu}$ and $\Gamma_{\hat{4}}
=i\gamma_{5}$ where $\left\{\gamma_{\mu}\right\}$ is a representation of the
Dirac algebra in four-dimensions and $\gamma_{5}=i\gamma_{0}\gamma_{1}\gamma_{2}
\gamma_{3}$ is the chirality operator.

In terms of the torsion
\begin{equation}
  T^{C}_{\phantom{C}AB} = \left(E^{M}_{\phantom{M}A}E^{N}_{\phantom{N}B}
    - E^{M}_{\phantom{M}B}E^{N}_{\phantom{N}A}\right)\partial_{N}E_{M}^{\phantom{M}C},
\end{equation}
the spin connection is
\begin{equation}
  \omega_{M}^{\phantom{M}AB} = \frac{1}{2}E_{M}^{\phantom{M}C}
  \left(T_{C}^{\phantom{C}AB} - T^{AB}_{\phantom{AB}C}
    - T^{B\phantom{C}A}_{\phantom{B}C\phantom{A}}\right).
\end{equation}
The non-vanishing components of the spin connection are then
\begin{equation}
  \omega_{\mu}^{\phantom{\mu}\hat{4}b}=\frac{f'}{4f^{3/2}}\delta^{b}_{\mu}
\end{equation}
so that the covariant derivative is
\begin{align}
  D_{\mu} =& \partial_{\mu} - \frac{if'}{8f^{3/2}}\gamma_{\mu}\gamma_{5}
  \nonumber \\
  D_{r} =& \partial_{r}.
\end{align}
Defining $\hat{\Psi}=f^{-1/2}\Psi$ simplifies the action to
\begin{equation}
  S = \int d^{5}x\, f^{1/4}\left\{if^{1/4}\hat{\bar{\Psi}}\eta^{\mu\sigma}
    \gamma_{\mu}
    \partial_{\sigma}\hat{\Psi} - f^{-1/4}\hat{\bar{\Psi}}\gamma_{5}\partial_{r}
    \hat{\Psi} - M_{f}\hat{\bar{\Psi}}\hat{\Psi}\right\}.
\end{equation}
We can write
\begin{equation}
  \hat{\Psi} = \hat{\Psi}_{L} + \hat{\Psi}_{R},
\end{equation}
where $\gamma_{5}\hat{\Psi}_{L,R} = \mp\hat{\Psi}_{L,R}$, and perform the KK decomposition
\begin{equation}
  \hat{\Psi}_{L,R} = \sum_{n=0}^{\infty}\psi_{L,R}^{\left(n\right)}
  \left(x^{\mu}\right)
  \chi_{n}^{L,R}\left(r\right).
\end{equation}
Then, after integrating over $r$, the four-dimensional action becomes
(where all indices are now contracted with $\eta_{\mu\nu}$)
\begin{equation}
  S = \int d^{4}x\sum_{n=0}^{\infty}\left\{i\bar{\psi}_{L}^{\left(n\right)}
    \gamma^{\mu}\partial_{\mu}\psi_{L}^{\left(n\right)}
    + i\bar{\psi}_{R}^{\left(n\right)}\gamma^{\mu}\partial_{\mu}
    \psi_{R}^{\left(n\right)}
  - m_{n}\left(\bar{\psi}_{L}^{\left(n\right)}\psi_{R}^{\left(n\right)}
      + \bar{\psi}_{R}^{\left(n\right)}\psi_{L}^{\left(n\right)}\right)\right\},
\end{equation}
where we have imposed the orthonormality condition
\begin{equation}
  \label{eq:fermionnorm}
  \int d r\, f^{1/2}\chi_{n}^{L,R}\chi_{m}^{L,R}=\delta_{nm},
\end{equation}
and the internal wavefunctions satisfy the coupled equations
\begin{equation}
  \pm\partial_{r}\chi_{n}^{L,R} + f^{1/4}M_{f}\chi_{n}^{L,R}
  = f^{1/2}m_{n}\chi_{n}^{R,L}.
\end{equation}
Unlike the radial wave functions for the gauge bosons, these are first-order
equations.  We return to these equations in Sect.~4 for the modified gauge-fermion couplings with 5-d mass gaps.

\subsection{Spectra and couplings}

In the AdS background, the solutions to Eq.~\ref{eq:vkkeom} can be
written in terms of Bessel functions \cite{Davoudiasl:1999tf,Pomarol:1999ad}
\begin{equation}
  a_{n} = \frac{A_{n}}{r}J_{1}\left(\frac{m_{n}R^{2}}{r}\right)
  + \frac{B_{n}}{r}Y_{1}\left(\frac{m_{n}R^{2}}{r}\right).
\end{equation}
For the IR smooth backgrounds considered here, the solutions are more complex.
For the
special case in which $f_{2}=2\epsilon^{2}$, the solutions can be
expressed in terms of (analytically continued) associated Legendre functions
\begin{equation}
  a_{n} = \frac{1}{\bigl(r^{2} + r_{\mathrm{tip}}^{2}\bigr)^{1/4}}
  \left(\alpha_{\nu}P_{1/2}^{\nu}\left(\frac{ir}{r_{\mathrm{tip}}}\right)
    + \beta_{\nu}Q_{1/2}^{\nu}\left(\frac{ir}{r_{\mathrm{tip}}}\right)\right),
\end{equation}
where $\nu=\sqrt{\frac{1}{4}+\frac{m_{n}^{2}R^{4}}{r_{\mathrm{tip}}^{2}}}$.
Even though an analytic solution exists, the dependence on the KK mass
is highly nonlinear, entering through the order of the Legendre function.  When the warp factor cannot be expressed as a perfect square, $f(r) =
(R^{2}/(r_{\mathrm{tip}}^2 + r^2))^2$, analytic solutions are not found for
general $m_{n}$.
We instead solve the equation numerically.

We will consider scenarios in
which the gauge symmetries are broken by a Higgs vev and not by boundary
conditions.  Since a Dirichlet boundary condition explicitly breaks
gauge invariance, we apply Neumann conditions at each brane.  As for RS,
a flat zero mode solution with $m_{0}=0$ is admitted with these boundary
conditions.

As
expected, changing the geometry alters the spectrum.  When $R$ and
the hierarchy are left fixed, as $f_{2}$
increases, the masses of the KK modes increase while the spacing
between KK modes becomes tighter.  As was found for gravitons \cite{us}
the geometries give a distinguishable pattern from RS (Tables
\ref{table:photonmasses5d}-\ref{table:tevcoup10d},
Fig.~\ref{fig:f2kk}).  Like the KK graviton, the 10-d mass gap gives larger effects than the 5-d mass gap, with ${\cal O}(10)$ changes in the couplings to IR localized fields and ${\cal O}(1)$ changes in the ratio between the masses of the KK states (which depends only on the geometry and not on the overall scale).  Thus the KK gauge spectra, like the KK graviton spectra, can be a probe of the geometry of the warped compactification.
Changing the IR boundary condition to a Higgs boundary condition does not significantly change the KK masses for the values of $\tilde{v}R$
considered here.

\TABLE[ht]{
\begin{tabular}{|l||r|r|r|r|r||r|}
  \hline
  & \multicolumn{6}{c|}{$m_{n}$ (TeV)} \\
  \cline{2-7}
  Mode & $f_{2}=0$ & $f_{2}=\epsilon^{4}$ & $f_{2}=2\epsilon^{2.1}$
  & $f_{2}=2\epsilon^{2}$ & $f_{2}=2\epsilon^{1.9}$ & RS \\
  \hline\hline
  1 & 1.00 & 1.00 & 1.00 & 1.00 & 1.00 & 1.00 \\
  2 & 2.53 & 2.53 & 2.53 & 2.42 & 2.05 & 2.27 \\
  3 & 3.99 & 3.98 & 3.98 & 3.81 & 3.20 & 3.54 \\
  4 & 5.44 & 5.43 & 5.42 & 5.20 & 4.33 & 4.82 \\
  \hline
\end{tabular}
\caption{4-d masses for the first few modes in different 5-d geometries for
$r_{\mathrm{tip}}/R=\epsilon=4.2\times 10^{-16}$ with the first mode normalized to
the (unrealistic) value $1\mathrm{\ TeV}$.}
\label{table:photonmasses5d}
}

\TABLE[ht]{
\begin{tabular}{|l||r|r|r|r|r||r|}
  \hline
  & \multicolumn{6}{c|}{$g_{n}/g_{0}$} \\
  \cline{2-7}
  Mode & $f_{2}=0$ & $f_{2}=\epsilon^{4}$ & $f_{2}=2\epsilon^{2.1}$
  & $f_{2}=2\epsilon^{2}$ & $f_{2}=2\epsilon^{1.9}$ & RS \\
  \hline\hline
  0 & 1 & 1 & 1 & 1 & 1 & 1 \\
  1 & 6.95 & 6.96 & 6.97 & 7.34 & 9.85 & 8.39 \\
  2 & 6.24 & 6.24 & 6.24 & 6.88 & 10.7 & 8.41 \\
  3 & 6.27 & 6.27 & 6.27 & 6.82 & 10.4 & 8.43 \\
  4 & 6.27 & 6.27 & 6.27 & 6.80 & 10.3 & 8.43 \\
  \hline
\end{tabular}
\caption{Relative couplings of 5-d KK excitations to IR localized fields for
different geometries.  These are given by the ratios of wave function
values at the IR when the wavefunctions are
normalized in the sense of Eq.~\ref{eq:vnorm}.}
\label{table:tevcoup5d}
}

\TABLE[ht]{
\begin{tabular}{|l||r|r|r|r|r||r|}
  \hline
  & \multicolumn{6}{c|}{$m_{n}$ (TeV)} \\
  \cline{2-7}
  Mode & $f_{2}=0$ & $f_{2}=\epsilon^{4}$ & $f_{2}=2\epsilon^{2.1}$
  & $f_{2}=2\epsilon^{2}$ & $f_{2}=2\epsilon^{1.9}$ & RS \\
  \hline\hline
  1 & 1.00 & 1.00 & 1.00 & 1.00 & 1.00 & 1.00 \\
  2 & 1.67 & 1.67 & 1.74 & 1.64 & 1.31 & 2.27 \\
  3 & 2.29 & 2.29 & 2.36 & 2.24 & 1.62 & 3.54 \\
  4 & 2.90 & 2.90 & 3.02 & 2.83 & 1.94 & 4.82 \\
  \hline
\end{tabular}
\caption{4-d masses for the first few modes in different 10-d geometries
(with $S^{5}$ angular space) for
$r_{\mathrm{tip}}/R=\epsilon=4.2\times 10^{-16}$ with the first mode normalized to
$1\mathrm{\ TeV}$.}
\label{table:photonmasses10d}
}

\TABLE[ht]{
\begin{tabular}{|l||r|r|r|r|r||r|}
  \hline
  & \multicolumn{6}{c|}{$g_{n}/g_{0}$} \\
  \cline{2-7}
  Mode & $f_{2}=0$ & $f_{2}=\epsilon^{4}$ & $f_{2}=2\epsilon^{2.1}$
  & $f_{2}=2\epsilon^{2}$ & $f_{2}=2\epsilon^{1.9}$ & RS \\
  \hline\hline
  0 & 1 & 1 & 1 & 1 & 1 & 1 \\
  1 & 11.3 & 11.3 & 13.7 & 28.7 & 49.2 & 8.39 \\
  2 & 46.8 & 46.8 & 47.7 & 86.7 & 119 & 8.41 \\
  3 & 106 & 106 & 110 & 183 & 249 & 8.43 \\
  4 & 193 & 193 & 197 & 323 & 461 & 8.43 \\
  \hline
\end{tabular}
\caption{Relative couplings of 10-d KK excitations to IR localized fields for
different geometries when the angular space is $S^{5}$.
\label{table:tevcoup10d}}
}

\EPSFIGURE[ht]{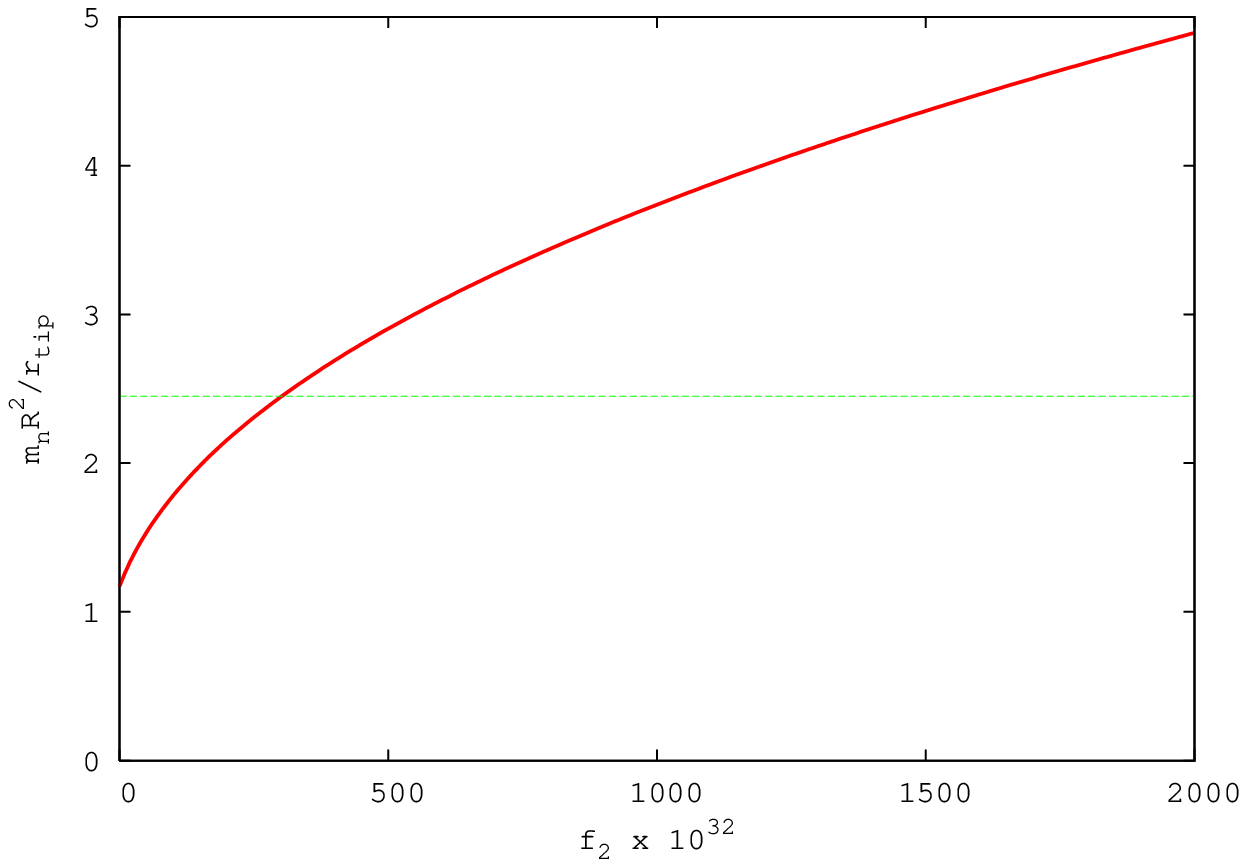,scale=.9}{Lowest KK mass in
  different geometries.  The value for RS is $2.45$, which is shown
  in green.~\label{fig:f2kk}}

Profiles for the first and second KK modes for the internal wavefunctions of
a gauge field are shown in Figs.~\ref{fig:mode1-5d}-\ref{fig:mode1-10d}.  As
$f_{2}$ increases, the wave function becomes increasingly localized towards the
IR when the wavefunctions are normalized in the sense of Eq.~\ref{eq:vnorm}.
This effect is enhanced in 10-d, leading to strong coupling between the TeV
fields and the KK modes.  Although the strength of the coupling is such that
the physics can no longer be treated perturbatively for larger KK modes, it
would guarantee strong collider signals.

\EPSFIGURE[ht]{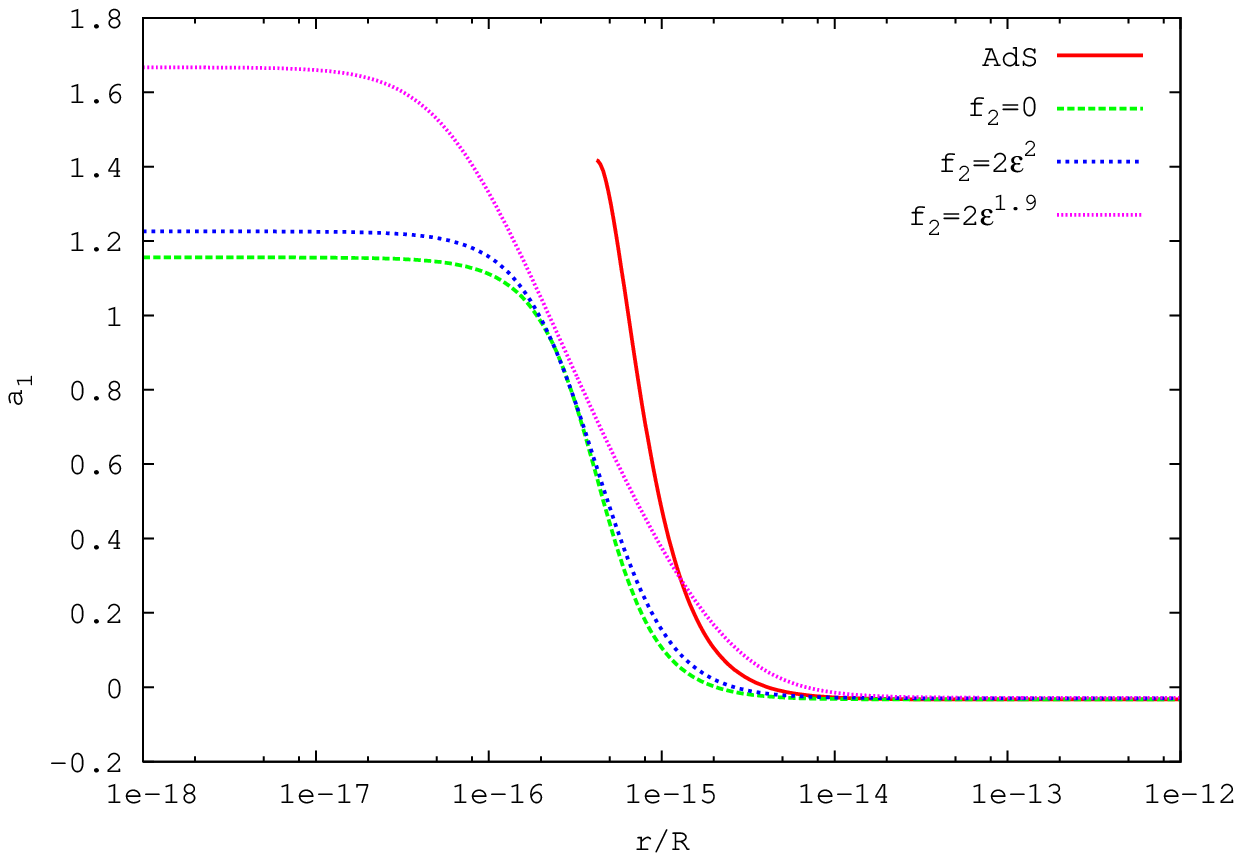,scale=.9}
{Internal wavefunctions for the first KK in 5-d mode in units of
$1/\sqrt{R}$.\label{fig:mode1-5d}}
\EPSFIGURE[ht]{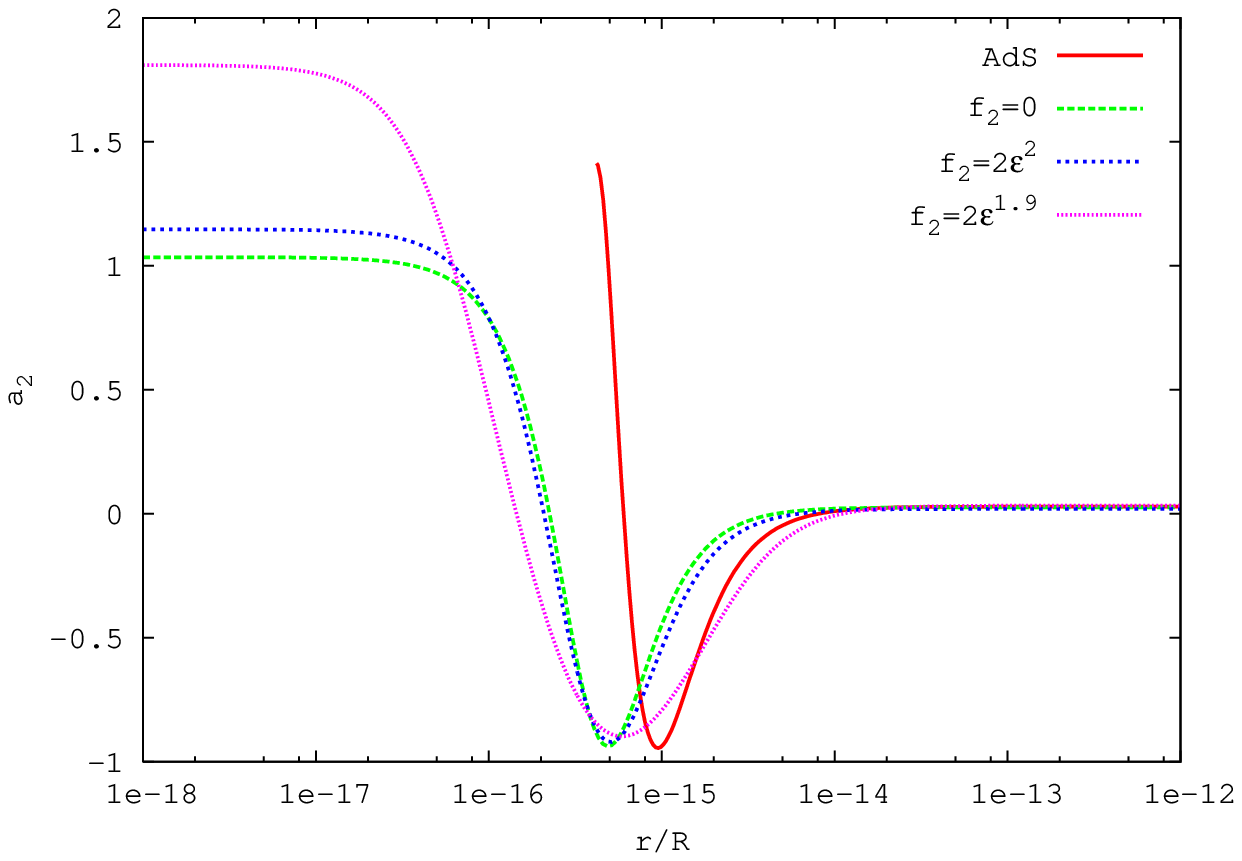,scale=.9}
{Internal wavefunctions for the second KK mode in 5-d in units of
$1/\sqrt{R}$.\label{fig:mode2-5d}}
\EPSFIGURE[ht]{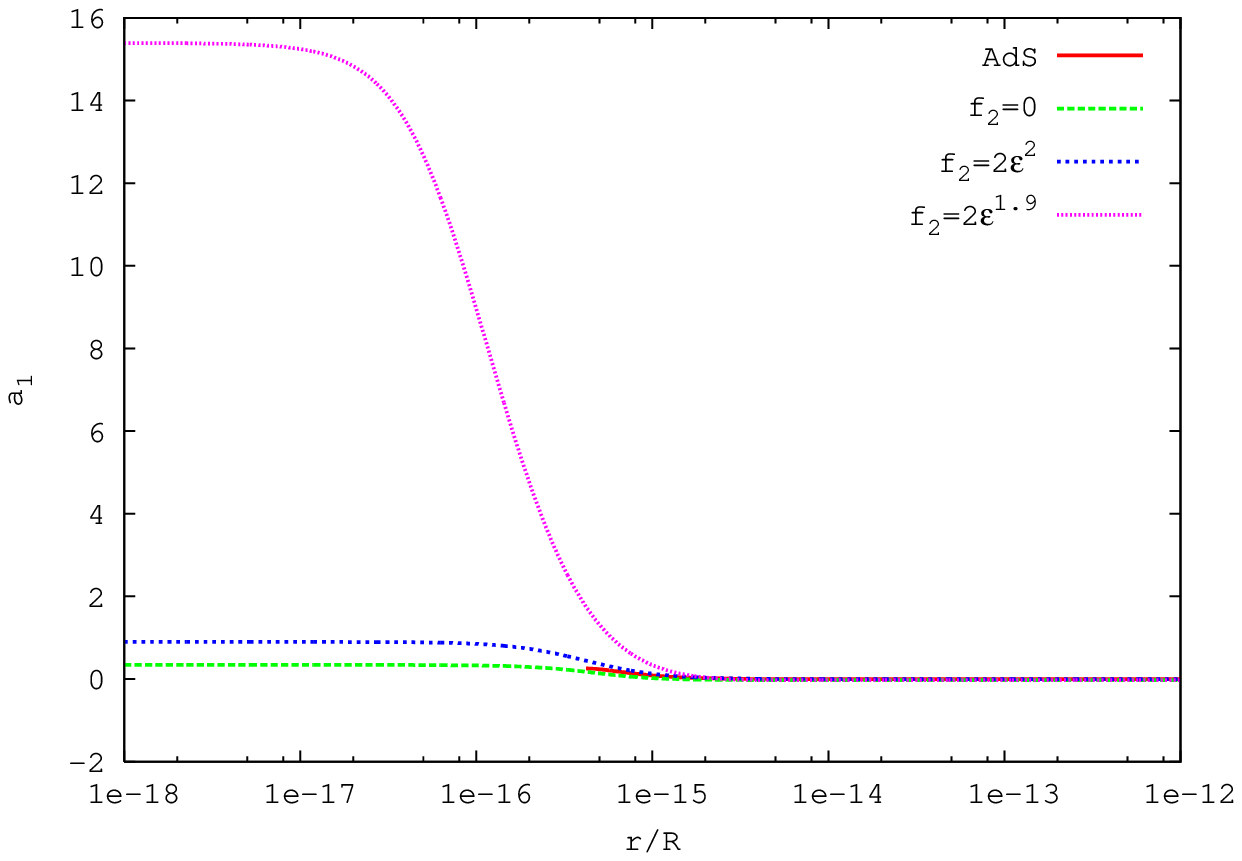,scale=.9}
{Internal wavefunctions for the first KK mode in 10-d in units of
$1/\sqrt{R}$.  The modes are more strongly localized to the
IR than they are in 5-d, leading to stronger couplings to IR localized fields.\label{fig:mode1-10d}}

The standard model fermions are associated with the lowest modes of
$\Psi_{L}$ while the zero mode of $\Psi_{R}$ is assumed to be projected out
by (for example) boundary conditions.  The bulk mass of $\Psi$ is written
$M_{f}=\nu/R$.  For $\nu\lesssim-\frac{1}{2}$, the profile  $\chi_{0}^{L}$ is localized
toward the IR brane, while for $\nu\gtrsim-\frac{1}{2}$, the profile is localized
toward the UV brane.  For the case in which a Higgs field that obtains a vev
is localized on the IR brane, the 4-d mass of the standard model fermion
is not zero, but is related to the overlap of the zero mode with the
Higgs.  More precisely, if $\Psi^{\left(\mathrm{d}\right)}$ transforms
as a part of a doublet under $SU(2)$ and $\Psi^{\left(\mathrm{s}\right)}$ is the
corresponding singlet, then the coupling to the Higgs vev is
\begin{equation}
  S = -\int d^{5}x\sqrt{G}\left(\lambda_{5}R\right)\frac{\tilde{v}}{2}
  \left(\bar{\Psi}
    ^{\left(\mathrm{d}\right)}\Psi^{\left(\mathrm{s}\right)} + \mathrm{h.c.}\right)
  \frac{\delta\left(r-r_{\mathrm{IR}}\right)}{\sqrt{G_{55}}}
\end{equation}
where $R$ is introduced to make $\lambda_{5}$ dimensionless (fermions are
dimension 2 in 5-d, while the Higgs vev is still dimension $1$ since the
Higgs is restricted to 4-d).  Using the delta function, this gives a 4-d
coupling to the warped Higgs vev $v=\tilde{v}\frac{r_{\mathrm{tip}}}{R}
\approx 246\ \mathrm{GeV}$
\begin{equation}
  \lambda_{4}=
  \lambda_{5}\frac{R^{2}}{r_{\mathrm{tip}}}\xi_{0}^{L\left(d\right)}
  \left(r=r_{\mathrm{IR}}\right)
  \xi_{0}^{L\left(s\right)}\left(r=r_{\mathrm{IR}}\right)
\end{equation}
Since the equation of motion is first-order, the value of the normalized radial
wave function on the IR (and hence the Yukawa coupling and the 4-d mass) is
determined by the values of $M_{f}$ for the singlet and the doublet. This
relation allows the hierarchy in Yukawa couplings to be explained by small
changes in the parameters $\nu$.  For
simplicity, we take the five dimensional Yukawa coupling $\lambda_{5}=1$ for every
fermion. Standard model fermions with small 4-d mass
have a weak coupling to the Higgs and are thus taken to be IR localized
($\nu\lesssim-\frac{1}{2}$) while heavier fermions are localized toward the UV
brane (Fig.~\ref{fig:leptonmass}).  Because the geometries considered here
differ from the RS scenario in only the IR, the physics for light
fermions is not significantly altered from RS in this scenario.  Since
the top has a large 4-d Yukawa coupling, either the left-handed or the
right-handed (singlet) top must be heavily localized towards the IR so we take $\nu$ for
the right-handed top to be equal to 1.  The value of $\nu$ for the
left-handed top is chosen such that the overlap with the Higgs gives the
correct 4-d top mass.  Since the left-handed top and the left-handed
bottom are in the same $SU\left(2\right)$ doublet, they must have the same
profile in the internal space (Fig.~\ref{fig:bprofile}).

\EPSFIGURE[ht]{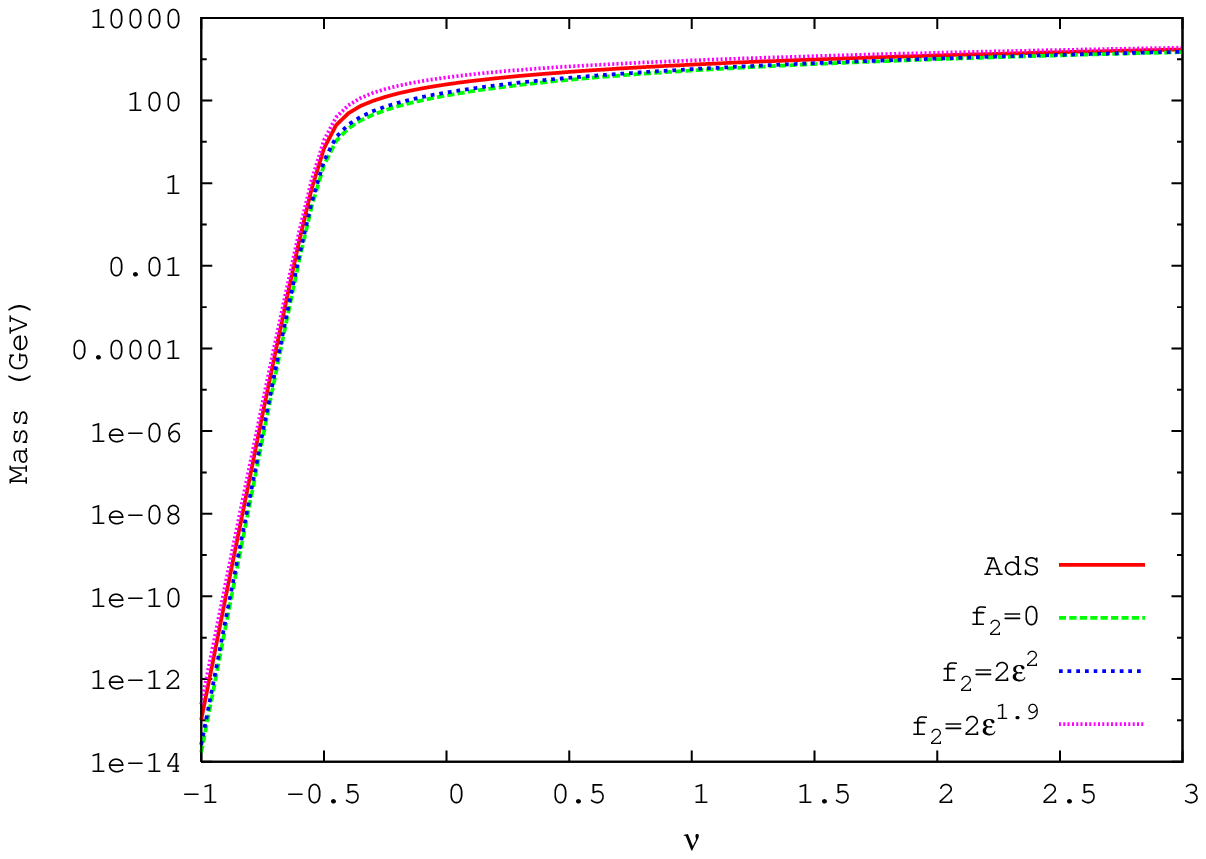,scale=.9}
{4-d masses for the lowest KK mode of $\Psi_{L}$  due to the presence of a Higgs
field localized on the IR brane assuming $\lambda_{5}=1$ and that $\nu$ is the
same for both the singlet and the doublet.
\label{fig:leptonmass}}

\EPSFIGURE[ht]{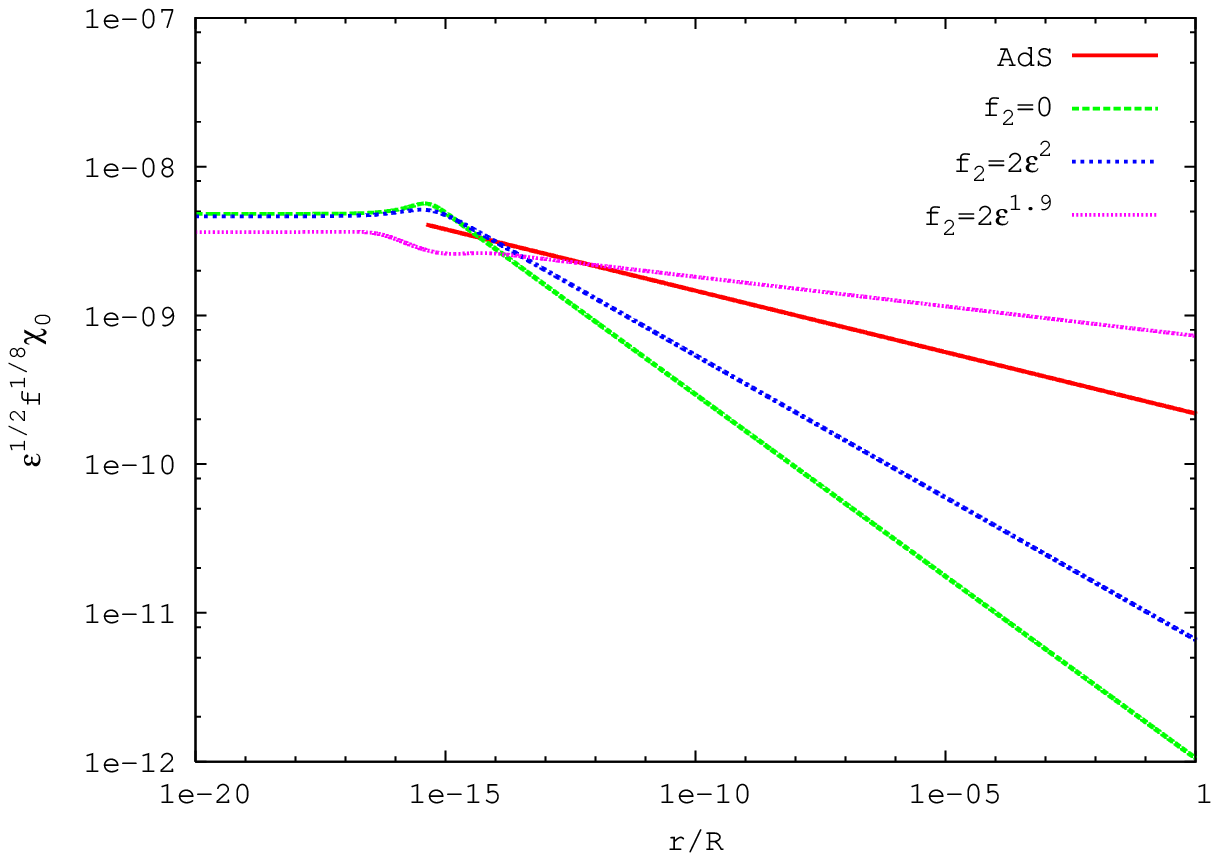,scale=.9}
{Internal wavefunction for the left-handed bottom quark in units of
$1/\sqrt{R}$, assuming $\lambda_{5}=1$ and that $\nu$ for the
right-handed top is $1$.  As $f_{2}$ decreases, the profile
becomes increasing localized in the IR.\label{fig:bprofile}}

\section{Precision electroweak constraints with 5-d mass gaps}

\label{PrecisionEW}

While allowing standard model fields, most notably the fermions, to propagate
in the bulk reduces some serious problems, such as flavor changing neutral
currents, it gives rise to other potential problems, such as
corrections to the precision electroweak observables $S$ and $T$ ($U$
remains
small for these models).  In the absence of a Higgs vev
on the TeV brane, Neumann b.c. may be consistently applied to both the UV and IR branes, and the wavefunctions of the zero modes of the gauge bosons are completely flat.  The presence of the Higgs vev on the IR brane forces the Neumann b.c. to be modified so that the derivative is non-vanishing there.  This modifies the gauge boson profiles from flatness in the extra dimension.  Since the W and Z gauge bosons have different couplings to the Higgs, their profiles in the extra dimension are modified in different ways.  The overlap of these profiles with the fermion profiles gives the gauge coupling of the gauge bosons to the fermions, which are now non-standard.   
For fermions localized in the same place in the extra dimension, the non-standard gauge couplings can be absorbed into a gauge boson wavefunction renormalization, and hence into the Peskin-Takeuchi parameters $S$ and $T$. In this class of models, the Peskin-Takeuchi parameters arise 
at tree-level, which we consider here.  For completeness, we describe this procedure in detail, following Csaki et al. \cite{Csaki1} though our normalization scheme differs from
theirs.

We begin by writing the KK decomposition for the
$SU\left(2\right)\times U\left(1\right)$ gauge fields
propagating in the bulk as
\begin{align}
  W_{\mu}^{\pm} =& \sum_{n=0}^{\infty}W_{\mu}^{\pm\left(n\right)}\left(x^{\mu}\right)
  w_{n}\left(r\right)  \notag \\
  Z_{\mu}=& \sum_{n=0}^{\infty}Z_{\mu}^{\left(n\right)}\left(x^{\mu}\right)
  z_{n}\left(r\right)  \notag \\
  A_{\mu} =& \sum_{n=0}^{\infty}A_{\mu}^{\left(n\right)}\left(x^{\mu}\right)
  a_{n}\left(r\right)  \notag.
\end{align}
The standard model fields are identified with $W^{\pm\left(0\right)}$,
$Z^{\left(0\right)}$, and $A^{\left(0\right)}$.  For the photon, which does not
couple to the Higgs, the corresponding internal wavefunction is a flat
true zero mode with zero 4-d mass.  However, the $W$ and $Z$ couple to
the IR-localized Higgs field, forcing the internal wavefunction to be
an almost zero mode and the 4-d mass to be non-vanishing.  In this
section,
we adopt a normalization for the internal wavefunctions different
from  Eq.~\ref{eq:vnorm} as explained below.

\subsection{Formalism}

The 5D action for the electroweak gauge fields (in a non-canonical
normalization) is
\begin{align}
  S =& \int d^{5}x\sqrt{G}\left\{-\frac{1}{4g_{5}^{2}}G^{MS}G^{MT}W_{MN}^{a}
      W_{ST}^{a} - \frac{1}{4g_{5}'^{2}}G^{MS}G^{NT}B_{MN}B_{ST}\right. \notag
    \\
    &+ \left.\frac{\tilde{v}^{2}}{8}\frac{\delta\left(r-r_{\mathrm{IR}}\right)}
      {\sqrt{G_{55}}}G^{MS}\left[W_{M}^{1}W_{S}^{1} + W_{M}^{2}W_{S}^{2}
      + \left(W_{M}^{3}-B_{M}\right)\left(W_{S}^{3}-B_{S}\right)\right]\right\}
\end{align}
where $W^{a}$ are the $SU\left(2\right)$ gauge fields and
$B$ is the $U\left(1\right)$ hypercharge gauge field.  In this
non-canonical normalization, the $Z$ and photon are written
\begin{align}
  Z_{M} =& W_{M}^{3} - B_{M} \\
  A_{M} =& s^{2}W_{M}^{3} + c^{2}Z_{M}
\end{align}
where the weak mixing angle is defined by the usual relation
\begin{equation}
  \sin\left(\theta_{\mathrm{W}}\right) = s = 
    \frac{g_{5}'}{\sqrt{g_{5}^{2}+g_{5}'^{2}}}~,
\end{equation}
and $c=\sqrt{1-s^2}$.

After integrating over the extra-dimension and keeping only the first member
of the KK towers, the action becomes
\begin{align}
  S =& \int d^{4}x\left\{-\frac{1}{2g^{2}}Z_{W}W_{\mu\nu}^{+}
    W^{-\mu\nu} - \frac{1}{4\left(g^{2} + g'^{2}\right)}Z_{Z}Z_{\mu\nu}Z^{\mu\nu}
    - \frac{1}{4e^{2}}Z_{\gamma}F_{\mu\nu}F^{\mu\nu}\right. \notag \\
  &+ \left. \left(\frac{v^{2}}{4} + \frac{1}{g^{2}}\Pi_{WW}\left(0\right)\right)
    W_{\mu}^{+}W^{-\mu}
    + \frac{1}{2}\left(\frac{v^{2}}{4} + \frac{1}{\left(g^{2}+g'^{2}\right)}
      \Pi_{ZZ}\left(0\right)\right)Z_{\mu}Z^{\mu}\right\},
\end{align}
where all indices have been contracted with $\eta_{\mu\nu}$ and
\begin{align}
  \frac{1}{g^{2}}Z_{W}=&\frac{1}{g_{5}^{2}}\int dr\, f^{1/4}w_{0}w_{0}
  \label{waverenorm}\\ \nonumber
  \frac{1}{g^{2} + g'^{2}}Z_{Z} =& \frac{1}{g_{5}^{2} + g_{5}'^{2}}\int dr\,
  f^{1/4}z_{0}z_{0} \\ \nonumber
  \frac{1}{e^{2}}Z_{\gamma} =& \left(\frac{1}{g_{5}^{2}} + \frac{1}{g_{5}'^{2}}
    \right)
  \int dr\,f^{1/4}a_{0}a_{0} 
  \label{Zs}
\end{align}
and
\begin{align}
  \frac{1}{g^{2}}\Pi_{WW}\left(0\right) =& \frac{1}{g_{5}^{2}}\int d r\,
  f^{-3/4}\left(\partial_{r}w_{0}\right)^{2} + \frac{v^{2}}{4}\bigl(
    w_{0}\left(r_{\mathrm{IR}}\right)^{2}-1\bigr)\\ \nonumber
  \frac{1}{g^{2} + g'^{2}}\Pi_{ZZ}\left(0\right) =& \frac{1}{g_{5}^{2}
    + g_{5}'^{2}}\int d r\, f^{-3/4}\left(\partial_{r}z_{0}\right)^{2}
  + \frac{v^{2}}{4}\bigl(z_{0}\left(r_{\mathrm{IR}}\right)^{2}-1\bigr),
\label{massrenorm}
\end{align}
while $v$ is simply the warped Higgs vev
\begin{equation}
  v = \tilde{v}f^{-1/4} \approx 246\mathrm{\ GeV}.
\end{equation}
Here, $z_{0}, w_{0}$ are the almost zero modes for the $Z$ and $W^{\pm}$
fields while $a_{0}$ is the true zero mode for the photon.

The physical meaning of these choices can be understood easily.  In the
absence of the Higgs vev, the wavefunctions for the gauge bosons are all
constants and equal.  When this is the case, we may take $Z_{X}=1$ and then
Eqs.~\ref{waverenorm} are a dependent set of equations that simply relate the
5-d coupling constants to the 4-d coupling constants,
\begin{equation}
  \frac{1}{g_{x}}=\frac{1}{g_{x5}^{2}}a^{2}\int dr\, f^{1/4}
\end{equation}
where $a=a_{0}\left(r\right)=w_{0}\left(r\right)=z_{0}\left(r\right)$ , $g_{x}$
is any of the 4-d couplings and $g_{x5}$ is the corresponding 5-d coupling. 
However,
when the Higgs has a non-zero vev, the wave functions for $W$ and $Z$ become modified
in the IR and Eqs.~\ref{waverenorm} can be satisfied only if some of
the $Z_{X}$ are different from unity, which results in corrections to
electroweak precision observables.

This can also be understood in the canonical normalization where the coupling
appears in the fermion terms.  As discussed below, the 4-d effective
coupling of the gauge fields to fermions can be written
\begin{align}
\label{canonical}
  g&= g_{5}\int dr\, f^{1/2}w_{0}\chi^{0}\chi^{0} \\
  \sqrt{g^{2}+g'^{2}} &= \sqrt{g_{5}^{2}+g_{5}'^{2}}\int dr\, f^{1/2}z_{0}
  \chi^{0}\chi^{0} \notag \\
  e = \frac{gg'}{\sqrt{g^{2}+g'^{2}}}=&\frac{g_{5}g_{5}'}{\sqrt{g_{5}^{2}
      + g_{5}'^{2}}}\int dr\, f^{1/2}w_{0}\chi^{0}\chi^{0} \notag
\end{align}
When the wavefunctions are flat, these equations are again a redundant set
of equations.  However, when the wavefunctions for the $Z$ and $W$ are
modified by the Higgs vev, these equations can only be satisfied if the
normalization of the wavefunctions for the heavy gauge bosons are modified,
again giving rise to corrections to the well measured properties of the $W$ and $Z$.

We normalize the wavefunctions so that $a_{0}=w_{0}=z_{0}=1$ on the UV brane.
Since light fermions must have a small overlap with the Higgs vev, they are
highly peaked at the
UV brane and this choice ensures that the coupling of these fermions to the
gauge bosons are as in the standard model.
This choice of normalization then redefines the corrections in precision EW observables into a wavefunction and mass renormalizations, Eqs.~\ref{waverenorm}, appearing in $S$, $T$ and $U$.  The zero mode of the photon,
which does not couple to the Higgs and is thus a true zero mode takes the
constant value $a_{0}=1$.

Given the discussion above, it is natural to choose $Z_{\gamma}=1$ since the
photon remains a true zero mode, in which case
\begin{equation}
  \frac{1}{e^{2}} = \left(\frac{1}{g_{5}^{2}} + \frac{1}{g_{5}'^{2}}\right)
  \int dr\, f^{1/4}.
\end{equation}
Then in order to preserve the relations between the coupling constants and
the weak mixing angles, we must take
\begin{align}
  \frac{1}{g^{2}} =& \frac{1}{g_{5}^{2}}\int dr\, f^{1/4} \\
  \frac{1}{g^{2}+g'^{2}} =& \frac{1}{g_{5}^{2} + g_{5}'^{2}}\int dr\, f^{1/4}.
\end{align}
Then the wavefunction renormalizations and mass corrections become
\begin{align}
  Z_{W} =& \frac{\int dr\, f^{1/4}w_{0}w_{0}}{\int dr\, f^{1/4}} \\
  Z_{Z} =& \frac{\int dr\, f^{1/4}z_{0}z_{0}}{\int dr\, f^{1/4}} \\
  \Pi_{WW}\left(0\right) =& \frac{\int dr\, f^{-3/4}\left(\partial_{r}w_{0}
    \right)^{2}}{\int dr\, f^{1/4}} + \frac{g^{2}v^{2}}{4}\bigl(
  w_{0}\left(r_{\mathrm{IR}}\right)^{2}-1\bigr) \\
  \Pi_{ZZ}\left(0\right) =& \frac{\int dr\, f^{-3/4}\left(\partial_{r}z_{0}
    \right)^{2}}{\int dr\, f^{1/4}} + \frac{\left(g^{2}+g'^{2}\right)v^{2}}{4}
  \bigl(z_{0}\left(r_{\mathrm{IR}}\right)^{2}-1\bigr)
\end{align}
We can relate these quantities to the oblique parameters $S, T,\text{and }U$. 
We identify
\begin{align}
  Z_{W} =& 1-g^{2}\Pi_{11}' \\
  Z_{Z} =& 1-\left(g^{2} + g'^{2}\right)\Pi_{33}' \\
  \Pi_{WW}\left(0\right) =& g^{2}\Pi_{11}\left(0\right) \\
  \Pi_{ZZ}\left(0\right) =& \left(g^{2} + g'^{2}\right)\Pi_{33}\left(0\right)
\end{align}
from which the Peskin-Takeuchi parameters\cite{Peskin:1990zt} are:
\begin{align}
  S =& 16\pi\Pi_{33}' \\
  T =& \frac{4\pi}{s^{2}c^{2}M_{Z}^{2}}\left(\Pi_{11}\left(0\right) 
    -\Pi_{33}\left(0\right)\right)\\
  U =& 16\pi\left(\Pi_{11}'-\Pi_{33}'\right) \\
\end{align}
Note that the expression for $S$ has simplified because of the lack of
$Z\!\!-\!\!\gamma$ mixing at the classical level.

In this context, the Peskin-Takeuchi parameters are a measure of how much
the wavefunctions of the $W$ and $Z$ deviate from a true zero mode in the
extra dimension.  When the Higgs is turned off, the full electroweak
symmetry is unbroken, the boundary conditions on the IR are restored to
Neumann boundary conditions and the Peskin-Takeuchi parameters vanish at
tree-level.

In the Randall-Sundrum scenarios, this deviation from the zero flat mode
leads to a large $T$ parameter unless the size of the extra dimension is
made
small.  This translates into a bound on the KK mass of order
10 TeV \cite{Csaki1}.  It is known however, that by introducing a custodial
$SU\left(2\right)$ symmetry into the bulk which is broken partially by
boundary conditions that $T$ can be made small.  The primary constraint on
$m_{KK}$ ($\gtrsim 3 \mbox{ TeV}$) is then derived from $S$ \cite{AgasheCustodial}.  Since the geometries give rise to significant shifts in the spectrum, one might suspect that constraints from $S$ and $T$ on the KK scale might be alleviated within these modified classes of metrics.  It turns out that this is not the case with the 5-d mass gap metrics, as we discuss next.

\subsection{Results}

The effect of the size (given in terms of
$R'=R/\epsilon$) and shape of the internal manifold on
$S$ and $T$
when no custodial symmetry is implemented is shown in Figs.~\ref{fig:s-rp}
and~\ref{fig:t-rp}
($U$ remains negligibly small in this setup).  For
every value of $f_{2}$, $S$ and $T$ are
always positive and increase monotonically as $R$ increases.

\EPSFIGURE[ht]{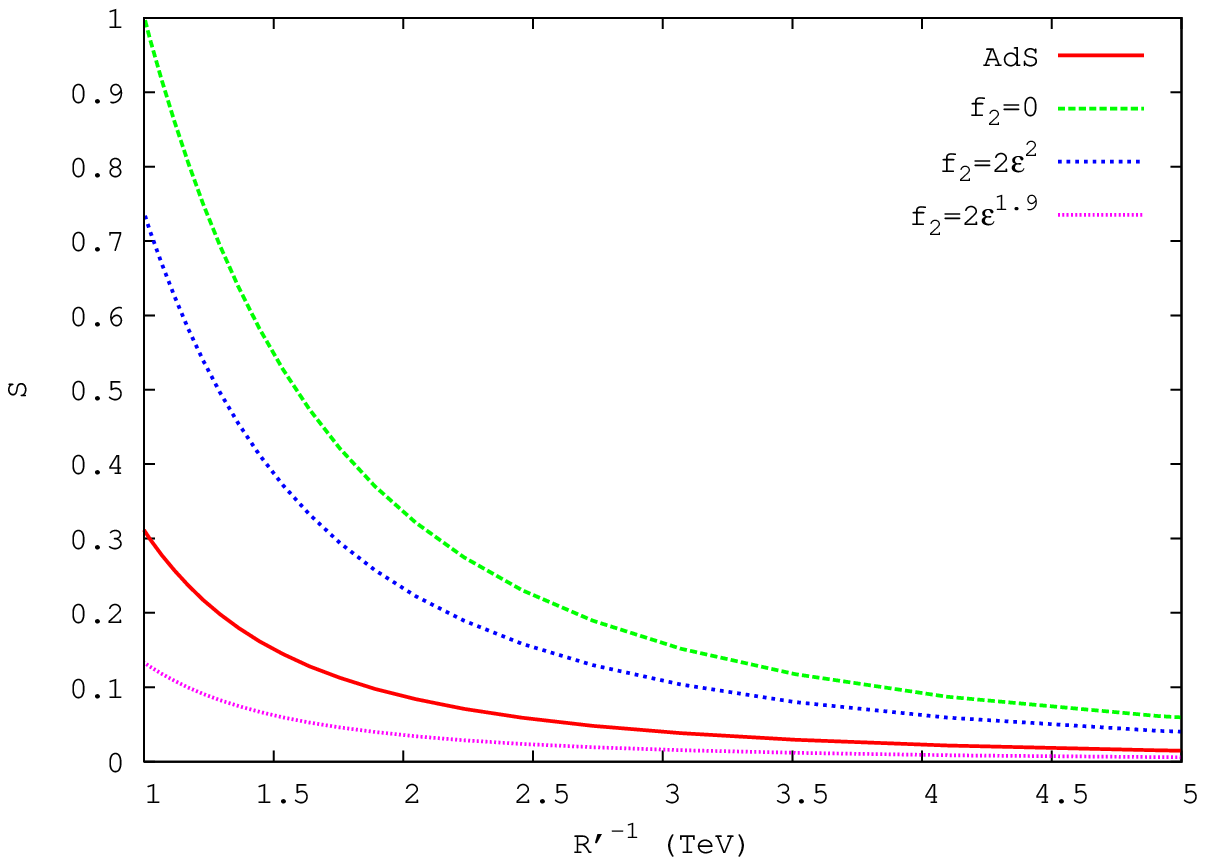,scale=.9}{The Peskin-Takeuchi
  parameter $S$ for different geometries where the hierarchy is fixed to
  $r_{\mathrm{tip}}/R=4.2\times 10^{-16}$.\label{fig:s-rp}}
\EPSFIGURE[ht]{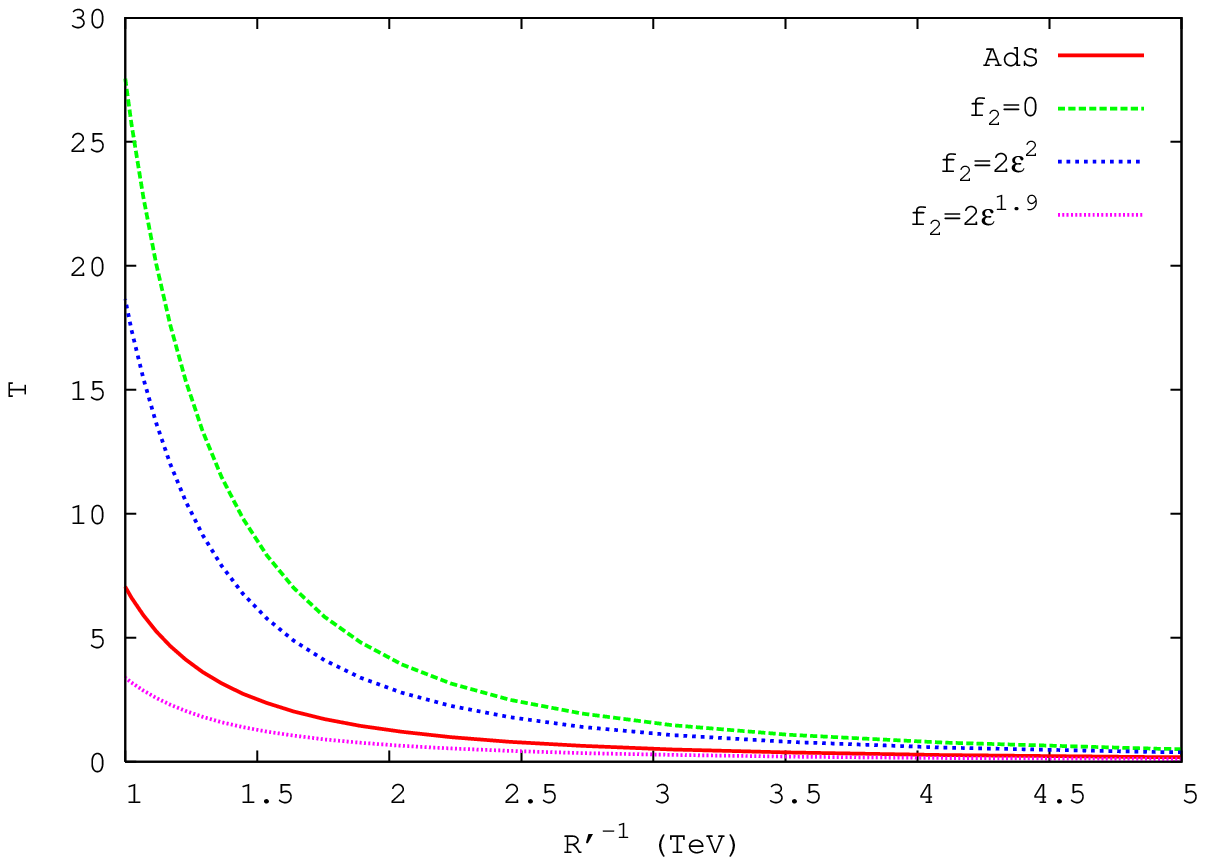,scale=.9}{The Peskin-Takeuchi
  parameter $T$ for different geometries where the hierarchy is fixed to
  $r_{\mathrm{tip}}/R=4.2\times 10^{-16}$. \label{fig:t-rp}}

Measurements on the $Z$ pole provide bounds on $S$ and $T$ and so
restrict the size of the extra dimensions and the mass of the first
KK excitation ($m_{\mathrm{KK}}$).  Since the functional dependence of
the Peskin-Takeuchi parameters on $R$ depends on the
geometry (see Fig.~\ref{fig:s-rp}), this might
a priori suggest that the bound on $m_{\mathrm{KK}}$ might change from one
geometry to another.  However, $m_{\mathrm{KK}}R^{2}/r_{\mathrm{tip}}$ also
changes with the geometry (Fig.~\ref{fig:f2kk}) in such a way that the two
effects cancel for $S$
(Fig.~\ref{s-mkk}) while $T$ changes
from one geometry to another though the changes remain
small for 5d mass gap metrics (Fig.~\ref{t-mkk}).  From Fig.~\ref{zprofile}, this
insensitivity 
to the geometry for UV localized $e$ and $\mu$ might have been expected.  In the UV, the $W$, $Z$ wavefunctions change very little from geometry to geometry as long as the mass of the first KK mode is normalized to the same value for each geometry.  
For fields localized near the UV brane, the effects of the KK modes can be summarized by their masses and their corresponding wavefunctions at the position of the Planck brane. 
With a fixed $m_{KK}$, the couplings of $W$ and $Z$ to $e$ and $\mu$ are not significantly affected by the IR modifications.  It is these couplings to $W$ and $Z$ in Eq.~\ref{canonical} , which are redefined into $S$ and $T$ through Eq.~\ref{Zs}.
If $e$ and $\mu$ were IR localized, one would expect a much bigger effect on $S$.
The slight variation of $T$ with the geometry is due to the fact that $T$ measures the derivatives of the 
$W$ and $Z$ wavefunctions, which are most significant in the IR (see Fig.~\ref{zprofile}) and hence the IR modifications can give rise to a shift in $T$.   
Our results are consistent with those of \cite{Delgado}, who found that $S$ and $T$ depend on volume factors, which differ little in the geometries we consider here.  In the future, we plan to consider the effects on $S$ and $T$ from 10-d mass gap geometries where 
the shifts on the Peskin-Takeuchi parameters (in particular, T) can be more significant.

\EPSFIGURE[ht]{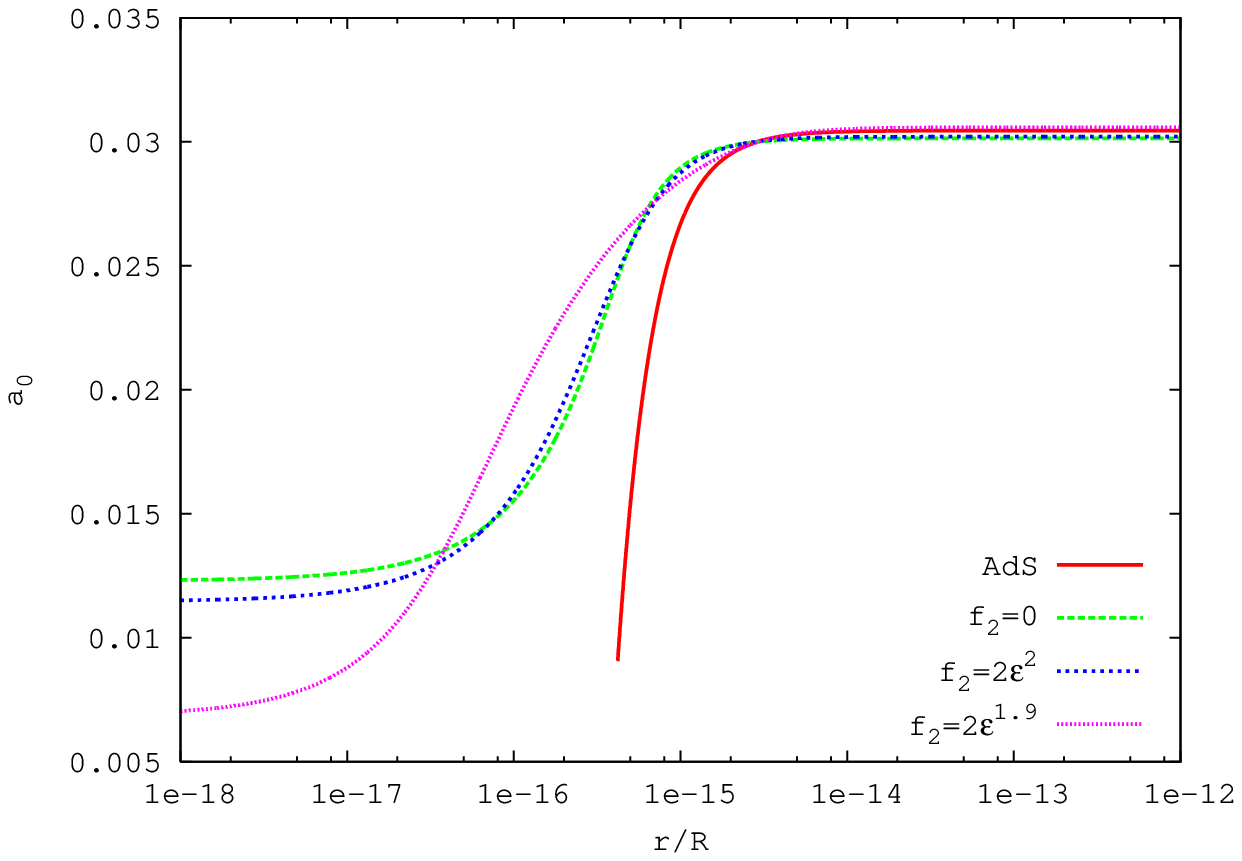,scale=.95}{The ``almost zero'' internal wavefunctions for
the $Z$ given in units of $1/\sqrt{R}$ (normalized in the sense of
eq.~\ref{eq:vnorm}).  The deviation from flatness is due to
the presence of the Higgs vev on the IR.  For each geometry, $R$ is chosen such
that $m_{\mathrm{KK}}=3\ \mathrm{TeV}$.\label{zprofile}}

In models that possess a custodial symmetry, the primary bound on
$m_{\mathrm{KK}}$ comes from $S$.  This implies that the bound on the
$m_{\mathrm{KK}}$ is independent of the shape of the internal space for the
warped geometries chosen here.  For models with custodial
symmetry, the bound on $m_{\mathrm{KK}}$ remains at $3-4$ TeV
\cite{AgasheCustodial}.

\EPSFIGURE[ht]{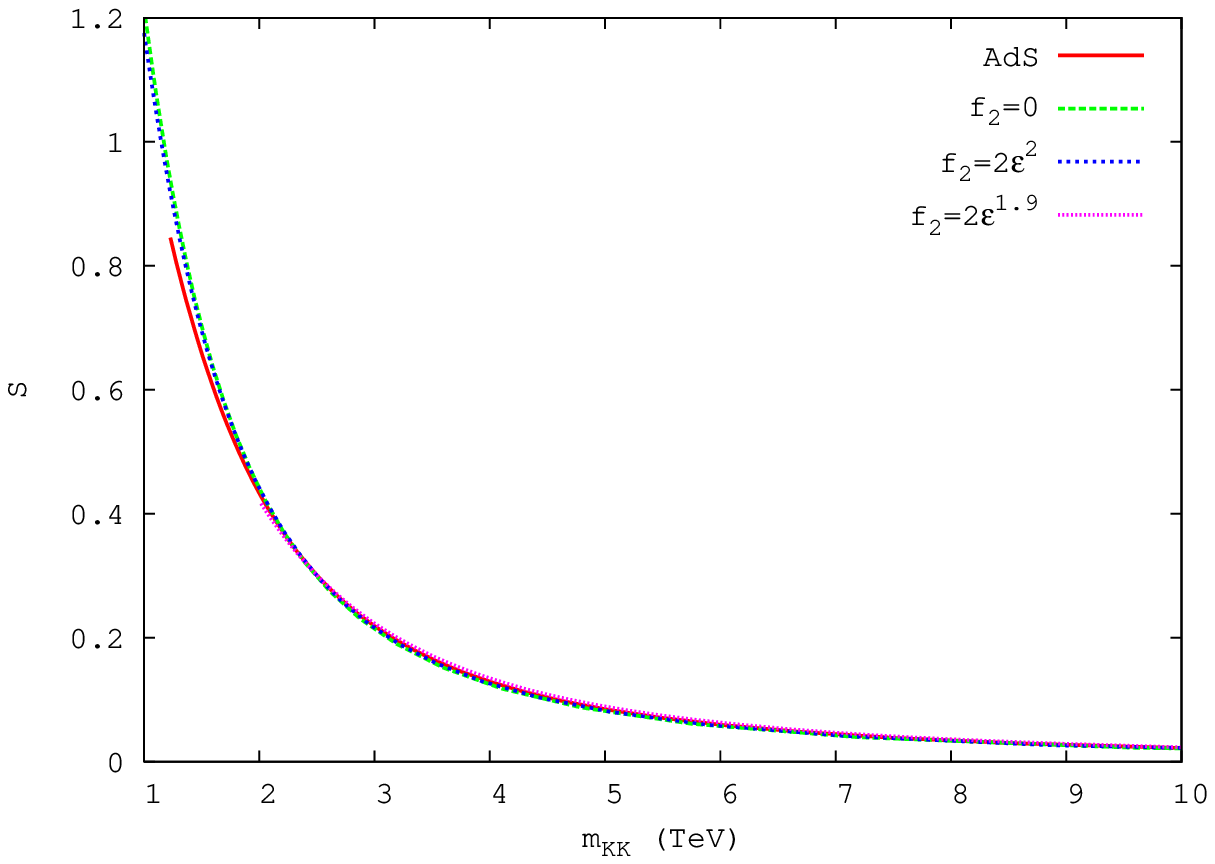,scale=.95}{The Peskin-Takeuchi
  parameter $S$ for different geometries and different values of
  $m_{\mathrm{KK}}$.  Since the curves lie essentially on top of each other, the constraint on $m_{KK}$ from $S$ remains unchanged.
  \label{s-mkk}}
\EPSFIGURE[ht]{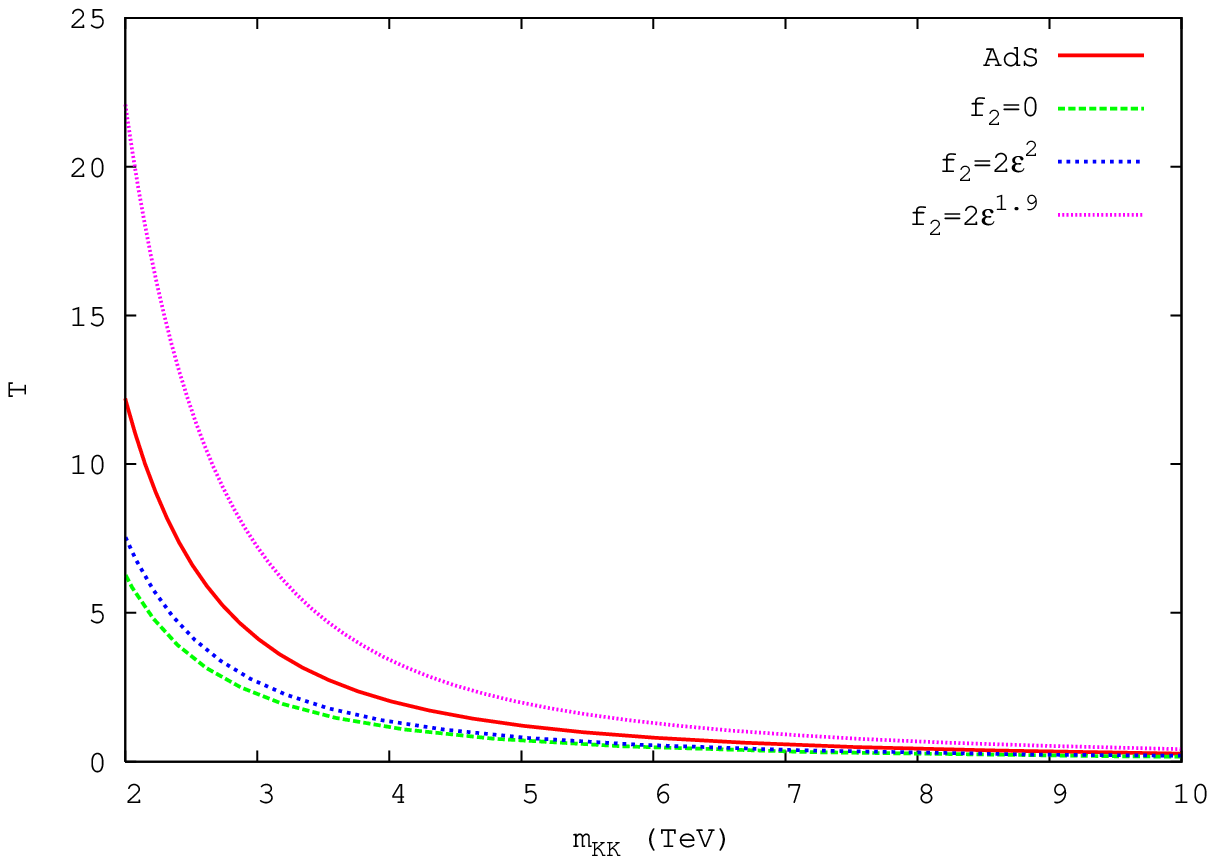,scale=.9}{The Peskin-Takeuchi
  parameter $T$ for different geometries and different values of
  $m_{\mathrm{KK}}$.\label{t-mkk}}

\section{Fermion-gauge boson couplings}

The action for a 5-d Dirac fermion coupling minimally to a 5-d gauge field with 
coupling constant $g_{5}$ is given by
\begin{equation}
  S=\int d^{5}x\sqrt{G}g_{5}\bar{\Psi}\Gamma^{A}E^{M}_{\phantom{M}A}A_{M}\Psi.
\end{equation}

With the gauge choice $A_{r}=0$, we find after performing the KK decomposition
and integrating over $r$,
\begin{equation}
  S = \int d^{4}x\sum_{n,m,q=0}^{\infty}\left\{
    g_{nm,q}^{L}\bar{\psi}_{L}^{\left(n\right)}\gamma^{\mu}
    A_{\mu}^{\left(q\right)}\psi_{L}^{\left(m\right)}
    + g_{nm,q}^{R}\bar{\psi}_{R}^{\left(n\right)}\gamma^{\mu}
    A_{\mu}^{\left(q\right)}\psi_{R}^{\left(m\right)}\right\},
\end{equation}
where the effective four-dimensional couplings are given by the overlap
\begin{equation}
  g_{nm,q}^{L,R}=g_{5}\int dr f^{1/2}\chi_{n}^{L,R}\chi_{m}^{L,R}a_{q}.
\end{equation}

Electroweak precision data come mostly from the physics of lighter fermions.
As discussed above and in \cite{Grossman:1999ra,Davoudiasl2}, lighter fermions have small coupling to the
Higgs and so must be localized towards the UV.  The localization is
determined by the mass $M=\nu/R$ such that when $\nu \lesssim -\frac{1}{2}$, the
wavefunction is localized towards the IR while for $\nu \gtrsim -\frac{1}{2}$,
the
wavefunction is localized towards the UV.  In order to render the
corrections to electroweak physics from the new physics oblique, we choose the
normalization of the gauge field so that the coupling of the gauge fields
to a standard model fermion localized entirely on the UV is equal to the
standard model value. This requires that $a_{0}=z_{0}=w_{0}=1$ on the UV brane.

The presence of the Higgs on the TeV brane alters the boundary condition of
the internal wavefunctions for the $W^{\pm}$ and the $Z$.  Since the 4-d
coupling constant is related to the overlap between the different KK
wavefunctions, this effects a change in the coupling between the fermion and
the boson.  For light
fermions which are localized towards the UV, this is a small effect since
the modification to the gauge boson wavefunction is restricted to the tip
in this normalization scheme (Fig.~\ref{fig:coupling}).  For fermions localized towards the IR, the geometry can significantly affect the couplings to the gauge fields.

\EPSFIGURE[ht]{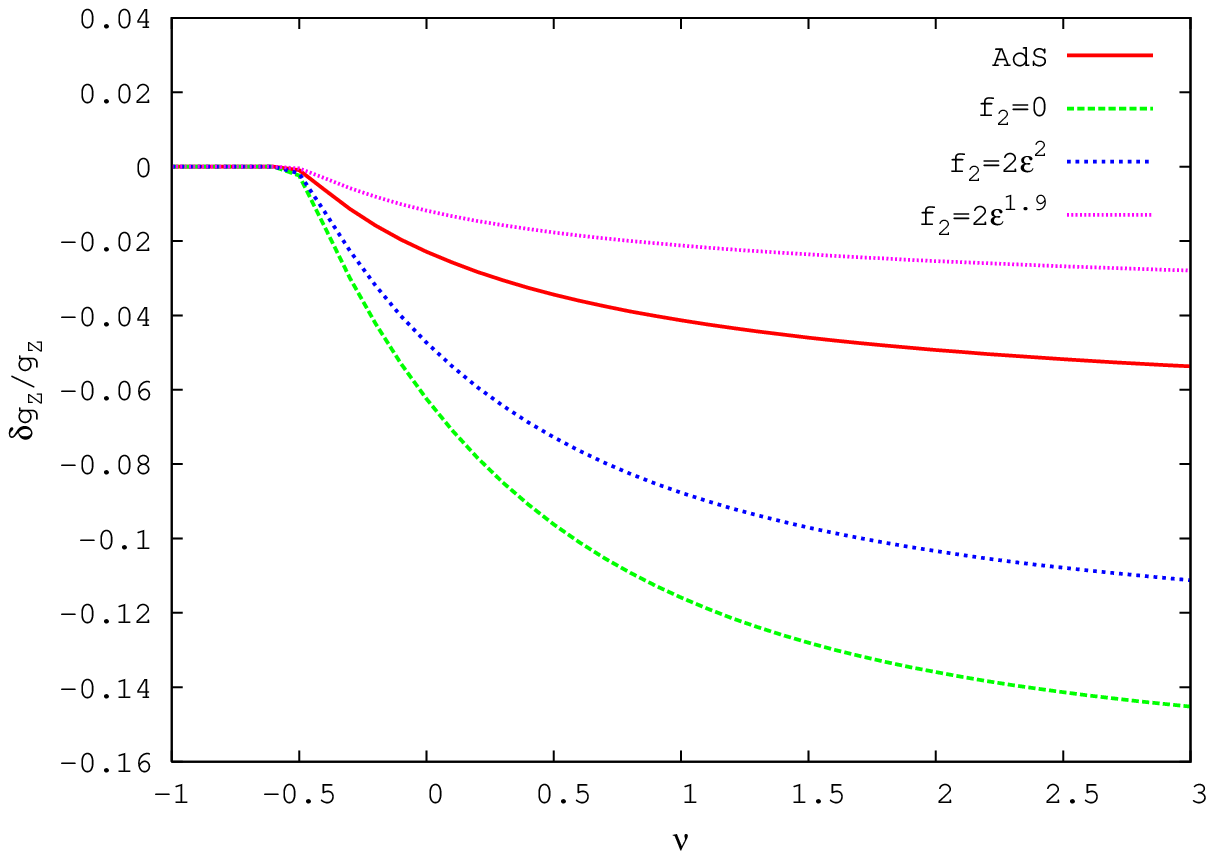,scale=.9}{Relative correction
  of the coupling of a fermion to the $Z$ almost zero mode due to the turning
  on of the Higgs vev.  $\nu=M_{f}R$ sets the location of the fermion
  such that for $\nu\lesssim -\frac{1}{2}$ the fermion is
  localized toward the UV while for $\nu\gtrsim \frac{1}{2}$ it is localized
  toward the IR.  Here, $R=3.4\times 10^{-16}\mathrm{\ TeV}^{-1}$ so that for
  RS, $m_{\mathrm{KK}}=3\mathrm{\ TeV}$.  Smaller values of $R$ naturally
  produce smaller effects.\label{fig:coupling}}

Precision measurements constrain the deviation of the coupling of the bottom
to the Z relative to the standard model value to be less than $\sim 1\%$.  Since
the left-handed top must be somewhat localized toward the IR to give a large
Yukawa, the left-handed bottom will also be localized toward the IR.  Since
this is where the $Z$ wavefunction deviates from flat, this will give
corrections to the coupling of the left-handed $B$ to the $Z$.
For a few geometries, this correction at the
$m_{\mathrm{KK}}=3\ \mathrm{TeV}$ bound is shown in Table~\ref{table:zbb}.  For
any particular
geometry, this bound on the coupling translates to a bound on the
value for $\nu$ for the left-handed bottom.  This is shown in Fig.~\ref{fig:nubound}.
For only small values of $f_{2}$ does this provide a stronger bound on the
geometry than $S$.

\TABLE[ht]{
  \begin{tabular}{|c||c|c|c|}
  \hline
  Geometry & $1/R'$ (TeV) & $\nu_{\mathrm{bottom}}$ &
  $\delta g_{Z}/G$ \\
  \hline\hline
  AdS & $1.2$ & $-.42$ & $-6.7\times 10^{-3}$ \\
  $f_{2}=0$ & $2.6$ & $-.26$ & $-1.2\times 10^{-2}$ \\
  $f_{2}=2\epsilon^{2}$ & $2.1$ & $-.31$ & $-1.0\times 10^{-2}$ \\
  $f_{2}=2\epsilon^{1.9}$ & $.76$ & $-.46$ & $-4.2\times 10^{-3}$ \\
  \hline
\end{tabular}
\caption{Relative correction of bottom coupling to the $Z$.  Here
  $R'=R^{2}/r_{\mathrm{tip}}$ and $\nu_{\mathrm{bottom}}=M_{f}R$ is the bulk mass
  required for the left-handed top and bottom doublet to reproduce the 4-d mass
  for the top when the 5-d Yukawa coupling $\lambda_{5}=1$ and $\nu$ for
  the right-handed top is equal to $1$.  The size of the geometry is such
  that $m_{\mathrm{KK}}=3\ \mathrm{TeV}$.}
\label{table:zbb}
}

\EPSFIGURE[ht]{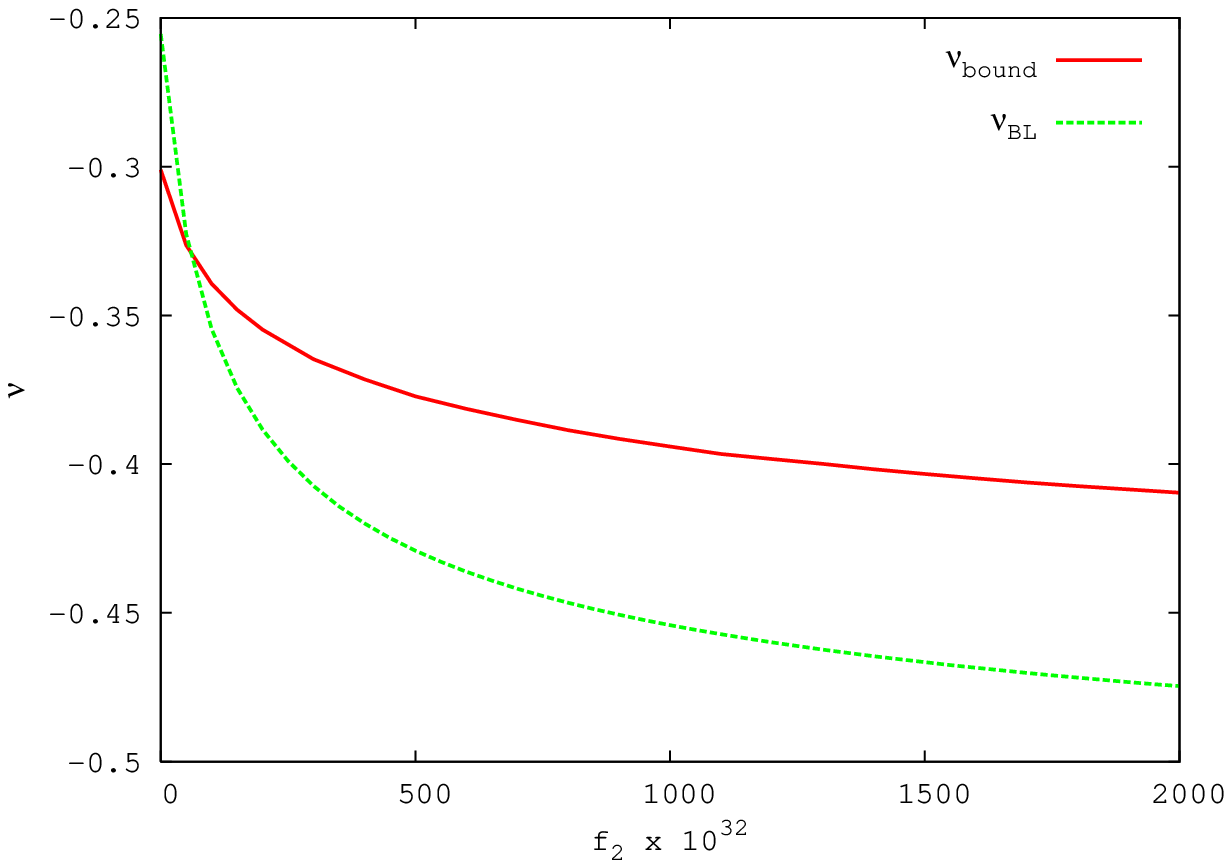,scale=.9}{Upper bound on
  $\nu$ consistent with $\delta g_{z}/g_{z}<1\%$ for different
  geometries.  The value of $\nu$ for the left-handed bottom is also
  shown.  This value is such that when $\lambda_{5}=1$ and $\nu$ for the
  right-handed top quark is $1$, the 4-d Yukawa coupling agrees with
  the observed top mass.\label{fig:nubound}}

The higher KK modes of the gauge bosons will also couple to the standard
model fermions.  These couplings again depend on the localization of the
fermions (Fig.~\ref{fig:coupling1}).

\EPSFIGURE[ht]{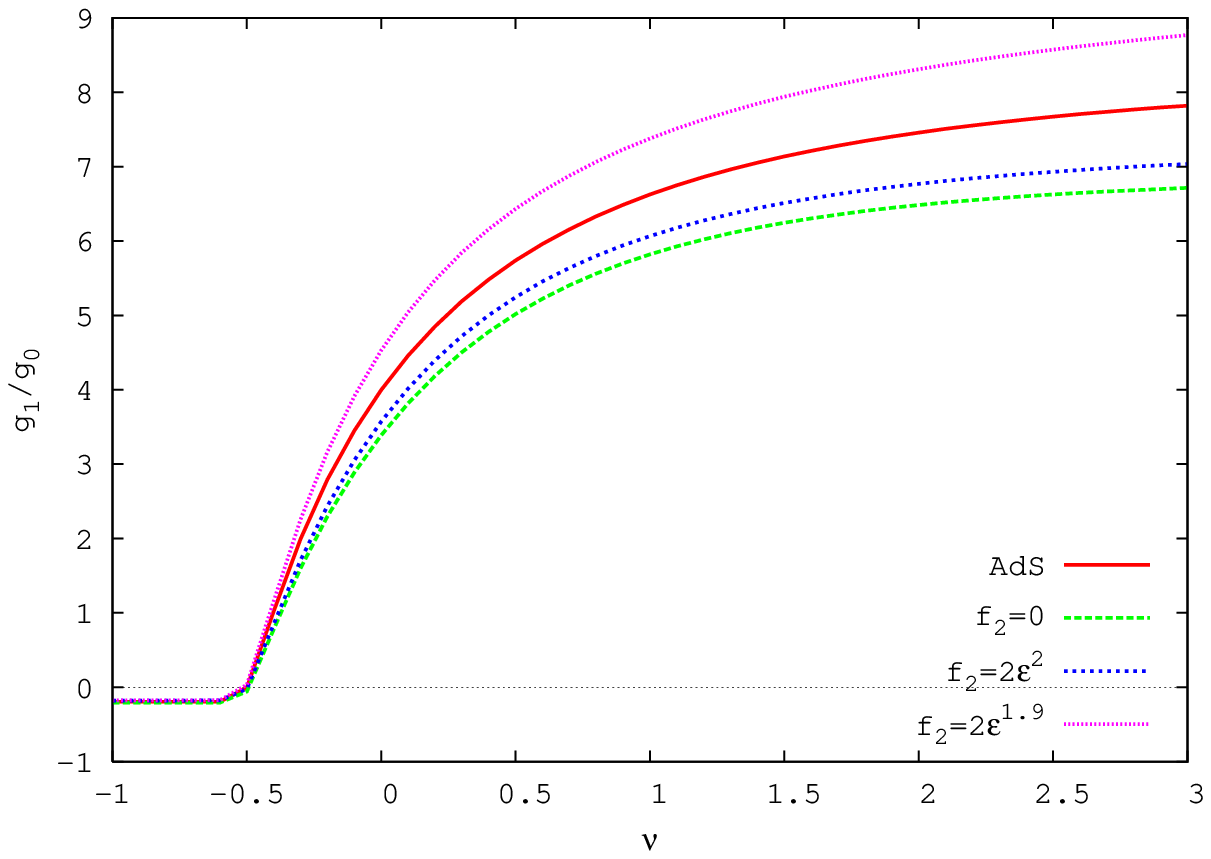,scale=.9}{Coupling of a bulk
  fermion to the first KK mode relative to the zeroth KK mode.  Because of
  the correction of the coupling to due a Higgs is small and the higher KK
  modes are almost entirely insensitive to the Higgs, this result applies
  whether or not the gauge boson couples to the Higgs vev.  As
  expected, the difference in geometries are more important for heavy
  (IR localized) fermions than for light (UV localized) fermions.
  \label{fig:coupling1}}

\section{Conclusions}

We have studied the effects of IR modifications of the AdS metric from a mass gap.  This metric, rather than becoming singular in the IR like AdS, levels off to a constant, breaking conformality.  We find that in both 5-d and 10-d, the KK gauge boson spectrum is sensitive to the form of the mass gap metric in the IR.  The 10-d mass gap metric yields the most significant changes, with factor $\sim 5-10$ increase in the coupling of the gauge KK fields to IR localized fields, and factor $\sim 2$ changes in the ratio of the KK masses.  With such large couplings to IR localized fields, even the lowest KK modes may be strongly coupled so that the standard perturbation theory analysis for production and decay no longer applies.   While these effects are quite large, especially in the 10-d case, they are not as large as those found for the KK graviton spectrum studied in our earlier paper \cite{us}.  This is expected, since the KK graviton is more highly peaked in the IR than the KK gauge
boson, so that it is more sensitive to IR modifications of the metric.  The shift of the top coupling to the $Z$, $\delta g_Z/g_z$ in 5-d, also depends on the IR geometry by a factor $\sim 2-4$.

For the precision EW analysis, the results of the standard analysis are basically robust against 5-d deformations of the metric in the IR.  There is {\em virtually no} change in $S$ with the new metrics.  $T$ is modified from AdS, but not so dramatically that the need for custodial symmetry can be removed.   We leave for future work an analysis of the effects of the 10-d metric on $S$ and $T$, since the result with the Higgs is sensitive to the way that the five extra compact dimensions are integrated out.

Remarkable advances in string phenomenology this decade has brought us closer to the ambitious goal of connecting string theory to data. 
Hopefully,
the LHC may allow us to probe not only the mechanism of electroweak symmetry breaking, but also
the geometry of string compactifications. The present work is a modest step towards this goal.

\section*{Acknowledgments}
It is a pleasure to thank Kaustubh Agashe, Ho Ling Li, Ben Lillie, Frank Petriello, and Bret
Underwood for
helpful discussions.
This work  was
supported in part by NSF CAREER Award No. PHY-0348093, DOE grant
DE-FG-02-95ER40896, a Research Innovation Award and a Cottrell
Scholar Award from Research Corporation.

\newpage

\bibliography{mg}

\end{document}